\documentclass[preprint,aps,showpacs,showkeys]{revtex4}
\usepackage{epsfig,amsmath,amssymb}
\renewcommand{\theequation}{\arabic{section}.\arabic{equation}}

\begin{document}

\title{Dynamic properties of the spin-1/2 $XY$ chain
       with three-site interactions}
\author{Taras Krokhmalskii$^{1,2}$,
        Oleg Derzhko$^{1,2}$,
        Joachim Stolze$^{2}$,
        and
        Taras Verkholyak$^{1,2}$}
\affiliation{$^1$Institute for Condensed Matter Physics,
             National Academy of Sciences of Ukraine,
             1 Svientsitskii Street, L'viv-11, 79011, Ukraine\\
             $^2$Institut f\"ur Physik, Technische Universit\"at Dortmund,
             44221, Dortmund, Germany}

\date{\today}

\begin{abstract}
We consider a spin-1/2 $XY$ chain in a transverse ($z$) field with multi-site interactions.
The additional terms introduced into the Hamiltonian 
involve products of spin components related to three adjacent sites.
A Jordan-Wigner transformation
leads to a simple bilinear Fermi form for the resulting Hamiltonian
and hence the spin model admits a rigorous analysis.
We point out the close relationships between several variants of the model 
which were discussed separately in previous studies.
The ground-state phases 
(ferromagnet and two kinds of spin liquid)
of the model 
are reflected in the dynamic structure factors of the spin chains,
which are the main focus in this study.
First we consider the $zz$ dynamic structure factor
reporting for this quantity a closed-form expression
and analyzing the properties of the two-fermion (particle-hole) excitation continuum
which governs the  dynamics of transverse spin component fluctuations
and of some other local operator fluctuations.
Then we examine the $xx$ dynamic structure factor
which is governed by many-fermion excitations, 
reporting both analytical and numerical results.
We discuss some easily recognized features of the dynamic structure factors
which are signatures for the presence of the three-site interactions.
\end{abstract}



\maketitle

\clearpage


\section{Multi-site interactions and Jordan-Wigner fermionization}
\label{sec1}
\setcounter{equation}{0}

Spin-1/2 $XY$ chains provide an excellent ground for studying various properties of quantum many-particle systems
since after performing the Jordan-Wigner transformation
these spin models can be reduced to systems of noninteracting spinless fermions \cite{lsm_k}.
One interesting issue which has emerged recently in the theory of quantum spin systems
is the study of effects of multi-site interspin interactions.
Such interactions may arise, e.g., in an effective spin model for the standard Hubbard model at half filling
in higher orders (beyond $t^2/U$) of the strong-coupling $t/U$ expansion \cite{spin_from_hubb}.
Another example is provided by quantum spin systems with energy currents \cite{antal,ogata,eisler}.
As early as 1971 \cite{suzuki}
M.~Suzuki proposed generalized one-dimensional $XY$ models with multi-site interactions,
allowing for rigorous analysis by the Jordan-Wigner fermionization approach.
An exactly integrable spin-1/2 $XXZ$ quantum spin chain with three-site interactions was suggested in Ref. \onlinecite{tsvelik}
(see also Ref. \onlinecite{frahm}).
An $XY$ version of this model was considered independently in Ref. \onlinecite{gr}.
Another version of the spin-1/2 $XY$ chain with three-site interactions was suggested in Refs. \onlinecite{drd,tj}.
Later on spin-1/2 $XY$ chains with three-site interactions were considered in a number of papers
concerning quantum phase transitions, transport properties and entanglement \cite{lou1,lou2,lou3,lou4,pachos,yang,zhang,lou5,yu}.
Recently,
in Ref. \onlinecite{z1} the spin-1/2 $XY$ chain with alternating three-site interaction has been introduced,
whereas in Ref. \onlinecite{z2} dynamic characteristics
of a few quantum spin chains with multi-site interactions have been discussed.
However,
an exhaustive study of the dynamic properties of spin-1/2 $XY$ chains with multi-site interactions,
similar to that for conventional $XY$ chains \cite{taylor,dks,dksm},
has not been performed yet.
With our paper we attempt to fill this gap.

In what follows we consider the Hamiltonian
\begin{eqnarray}
\label{1.01}
H=\sum_n
\left(
J_n\left(s_n^xs_{n+1}^x+s_n^ys_{n+1}^y\right)
+
D_n\left(s_n^xs_{n+1}^y-s_n^ys_{n+1}^x\right)
\right.
\nonumber\\
\left.
+
K_n\left(s_n^xs_{n+1}^zs_{n+2}^x+s_n^ys_{n+1}^zs_{n+2}^y\right)
+
E_n\left(s_n^xs_{n+1}^zs_{n+2}^y-s_n^ys_{n+1}^zs_{n+2}^x\right)
\right.
\nonumber\\
\left.
+\Omega_ns_n^z
\right).
\end{eqnarray}
Here $J_n$ and $D_n$ are the isotropic $XY$ (or $XX$) exchange interaction
and the $z$-component of the Dzyaloshinskii-Moriya interaction
between the neighboring sites $n$ and $n+1$,
respectively.
$K_n$ and $E_n$ are two types of three-site exchange interactions
introduced in Refs. \onlinecite{drd,tj} and in Ref. \onlinecite{gr}, respectively
(see also Ref. \onlinecite{z3} where the general Hamiltonian (\ref{1.01}) was introduced as well).
$\Omega_n$ is the transverse ($z$) external magnetic field at the site $n$.
The sum in Eq. (\ref{1.01}) runs over all $N$ lattice sites;
boundary conditions (open or periodic) are not important
for the quantities to be calculated in the thermodynamic limit $N\to\infty$.
In this study in most cases we restrict ourselves to homogeneous chains
with site-independent values of the interspin interaction constants and field,
i.e. $J_n=J$ etc.

We start by discussing the symmetry properties of the Hamiltonian (\ref{1.01})
in order to show the close relations
between the models
of Ref. \onlinecite{gr} (i.e. with the $XZY-YZX$ type of three-site interactions)
and
of Refs. \onlinecite{drd,tj} (i.e. with the $XZX+YZY$ type of three-site interactions)
which were not discussed before.
Consider a local spin rotation around the $z$ axis
\begin{eqnarray}
\label{1.02}
s_n^x\to \tilde{s}_n^x
=s_n^x\cos\phi_n + s_n^y\sin\phi_n,
\nonumber\\
s_n^y\to \tilde{s}_n^y
=-s_n^x\sin\phi_n + s_n^y\cos\phi_n,
\nonumber\\
s_n^z\to \tilde{s}_n^z
=s_n^z.
\end{eqnarray}
Under that transformation,
the parameters for the interspin interactions in the Hamiltonian (\ref{1.01}) are mapped as follows:
\begin{eqnarray}
\label{1.03}
J_n\to \tilde{J}_n
=J_n\cos\left(\phi_{n+1}-\phi_n\right)+D_n\sin\left(\phi_{n+1}-\phi_n\right),
\nonumber\\
D_n\to \tilde{D}_n
=-J_n\sin\left(\phi_{n+1}-\phi_n\right)+D_n\cos\left(\phi_{n+1}-\phi_n\right),
\nonumber\\
K_n\to \tilde{K}_n
=K_n\cos\left(\phi_{n+2}-\phi_n\right)+E_n\sin\left(\phi_{n+2}-\phi_n\right),
\nonumber\\
E_n\to \tilde{E}_n
=-K_n\sin\left(\phi_{n+2}-\phi_n\right)+E_n\cos\left(\phi_{n+2}-\phi_n\right).
\end{eqnarray}
This shows clearly that the rotations (\ref{1.02}) may be employed
to simplify the Hamiltonian (\ref{1.01}) by eliminating some of the interactions.
For example,
in a homogeneous chain we may achieve
$\tilde{D}=0$ by setting $\phi_{n+1}-\phi_{n}=\varphi$
with $\tan\varphi=D/J$ \cite{perk}.
The remaining coupling constants then are
$\tilde{J}={\rm{sgn}}(J)\sqrt{J^2+D^2}$,
$\tilde{K}=\left(\left(J^2-D^2\right)K+2JDE\right)/\left(J^2+D^2\right)$,
$\tilde{E}=\left(-2JDK+\left(J^2-D^2\right)E\right)/\left(J^2+D^2\right)$.

More interestingly,
using Eqs. (\ref{1.02}), (\ref{1.03})
we can eliminate from the Hamiltonian Eq. (\ref{1.01})
either of the three-site interactions, $K_n$ or $E_n$.
Introducing
$\theta_n=\phi_{n+2}-\phi_n$
we note that $\tilde{K}_n=0$
if $\tan\theta_n=-K_n/E_n$.
These $\theta_n$ values can be used for calculating the $\tilde{J}_n$ and $\tilde{D}_n$
by using
$\phi_{2m}-\phi_{2m-1}
=\phi_2-\phi_1+\sum_{j=1}^{m-1}\left(\theta_{2j}-\theta_{2j-1}\right)$
and
$\phi_{2m+1}-\phi_{2m}
=\theta_{2m-1}-\phi_{2m}+\phi_{2m-1}$;
the surviving three-site coupling is
$\tilde{E}_n={\rm{sgn}}(E_n)\sqrt{E_n^2+K_n^2}$.
If, on the other hand,
we put $\tan\theta_n=E_n/K_n$
we get $\tilde{E_n}=0$
whereas $\tilde{K}_n={\rm{sgn}}(K_n)\sqrt{K_n^2+E_n^2}$.
For the uniform chain,
on which we mainly focus in what follows,
$\theta_n=\theta$,
$\phi_{2m}-\phi_{2m-1}=\phi_2-\phi_1$,
$\phi_{2m+1}-\phi_{2m}=\theta-\phi_2+\phi_1$,
and assuming $\phi_2-\phi_1=\theta/2$
we get
$\tilde{J}=J\cos(\theta/2)+D\sin(\theta/2)$,
$\tilde{D}=-J\sin(\theta/2)+D\cos(\theta/2)$
and
either
$\tilde{K}=0$,
$\tilde{E}={\rm{sgn}}(E)\sqrt{K^2+E^2}$
if $\tan\theta=-K/E$
or
$\tilde{E}=0$,
$\tilde{K}={\rm{sgn}}(K)\sqrt{K^2+E^2}$
if $\tan\theta=E/K$.

To summarize this part,
we have shown 
that while studying effects of three-site interactions on the basis of the model (\ref{1.01})
it would be sufficient to consider the model (\ref{1.01})
with either nonzero parameters $J_n$, $D_n$, $K_n$ or $J_n$, $D_n$, $E_n$
since the most general case when all four types of interactions have nonzero values
can be reduced either to the former case or to the latter case.
One consequence of this
is that the model considered in Ref. \onlinecite{tj} can be reduced to the model considered in Ref. \onlinecite{gr}.
Namely,
starting from the model with $J\ne 0$, $K\ne 0$, $D=E=0$
and choosing $\phi_{n+2}-\phi_n=\pi/2$, $\phi_{n+1}-\phi_n=\pi/4$, $\phi_n=n\pi/4$
we arrive at the model with
$\tilde{J}=J/\sqrt{2}$, $\tilde{D}=-J/\sqrt{2}$, $\tilde{E}=-K$, $\tilde{K}=0$.
Vice versa,
starting from the model with $J\ne 0$, $D\ne 0$, $E\ne 0$, $K=0$
and performing the same transformation
we arrive at the model with
$\tilde{J}=\left(J+D\right)/\sqrt{2}$, $\tilde{D}=\left(-J+D\right)/\sqrt{2}$, $\tilde{K}=E$, $\tilde{E}=0$.
In our study of the dynamic properties of quantum spin chains with three-site interactions
we focus on the case $J\ne 0$, $K\ne 0$, $D=E=0$
leaving a detailed study of other cases for the future.

The peculiar nature of the three-site interactions studied here
becomes clear after performing the Jordan-Wigner transformation
\begin{eqnarray}
\label{1.04}
s_n^+=s_n^x+{\rm{i}}s_n^y=P_{n-1}c_n^{\dagger},
\;\;\;
s_n^-=s_n^x-{\rm{i}}s_n^y=P_{n-1}c_n,
\nonumber\\
c_n^{\dagger}=P_{n-1}s_n^+,
\;\;\;
c_n=P_{n-1}s_n^-,
\nonumber\\
P_m=\prod_{j=1}^m\left(1-2c_j^{\dagger}c_j\right)=\prod_{j=1}^m\left(-2s_j^z\right).
\end{eqnarray}
Plugging (\ref{1.04}) into (\ref{1.01}) we find
\begin{eqnarray}
\label{1.05}
H=\sum_{n}
\left(
\frac{J_n+{\rm{i}}D_n}{2}c_n^{\dagger}c_{n+1}+\frac{J_n-{\rm{i}}D_n}{2}c_{n+1}^{\dagger}c_n
\right.
\nonumber\\
\left.
-\frac{K_n+{\rm{i}}E_n}{4}c_n^{\dagger}c_{n+2}-\frac{K_n-{\rm{i}}E_n}{4}c_{n+2}^{\dagger}c_n
\right.
\nonumber\\
\left.
+\Omega_n\left(c_n^{\dagger}c_n-\frac{1}{2}\right)
\right),
\end{eqnarray}
i.e. the Hamiltonian of the spin model is a simple bilinear form in terms of spinless fermions.
For the uniform case it is convenient to employ periodic boundary conditions 
in the spin Hamiltonian (\ref{1.01}).
That leads either to periodic or antiperiodic boundary conditions 
in the fermion Hamiltonian (\ref{1.05}),
depending on whether the number of spinless fermions is even or odd.
Either the nearest-neighbor hopping integrals
or the next-nearest-neighbor hopping integrals
can be made real (or purely imaginary) by applying a gauge transformation,
which is the analog of Eqs. (\ref{1.02}), (\ref{1.03}) in the spinless-fermion picture.

In the uniform case we can diagonalize (\ref{1.05}) by performing the Fourier transformation
\begin{eqnarray}
\label{1.06}
c_n^{\dagger}=\frac{1}{\sqrt{N}}\sum_{\kappa}\exp\left({\rm{i}}\kappa n\right) c_\kappa^{\dagger},
\;\;\;
c_n=\frac{1}{\sqrt{N}}\sum_{\kappa}\exp\left(-{\rm{i}}\kappa n\right) c_\kappa,
\end{eqnarray}
with $\kappa=(2\pi/N)m$ [$\kappa=(2\pi/N)(m+1/2)$] in the subspaces with odd [even] numbers of spinless fermions
and $m=-N/2, -N/2+1, \ldots, N/2-1$ (if $N$ is even)
or $m=-(N-1)/2, -(N-1)/2+1, \ldots, (N-1)/2$ (if $N$ is odd)
arriving at
\begin{eqnarray}
\label{1.07}
H=\sum_{\kappa}\Lambda_{\kappa}\left(c_{\kappa}^{\dagger}c_{\kappa}-\frac{1}{2}\right),
\nonumber\\
\Lambda_{\kappa}=J\cos\kappa+D\sin\kappa
-\frac{K}{2}\cos(2\kappa)-\frac{E}{2}\sin(2\kappa)
+\Omega.
\end{eqnarray}

\begin{figure}
\epsfig{file = 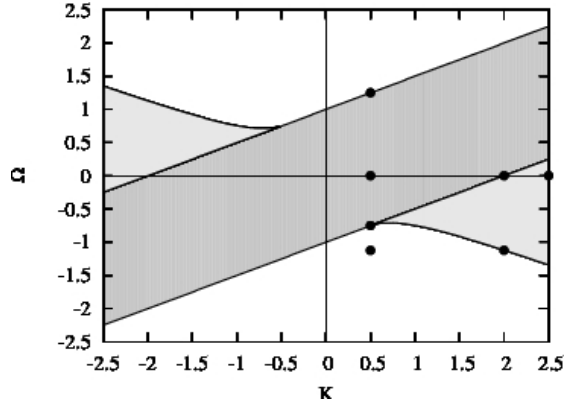, height = 0.32\linewidth}\\
\caption{The ground-state phase diagram of the homogeneous model (\ref{1.01})
with $J=\pm 1$, $K\ne 0$, $D=E=0$ discussed earlier in Ref. \onlinecite{tj}.
The region $-1+K/2<\Omega<1+K/2$ corresponds to the spin liquid I phase
(two Fermi points) [dark-gray],
the regions
$K<-1/2$, $1+K/2<\Omega<-K/2-1/(4K)$
and
$K>1/2$, $-K/2-1/(4K)<\Omega<-1+K/2$
correspond to the spin liquid II phase
(four Fermi points) [light-gray],
the remaining regions correspond to the ferromagnetic phase [light].
The black dots correspond to the sets of parameters
which we most often use below to discuss dynamic quantities.}
\label{fig01}
\end{figure}

From Eq. (\ref{1.07}) we immediately conclude
that the external magnetic field $\Omega$ plays the role of a chemical potential for spinless fermions.
More interestingly,
the two terms proportional to $\cos(2\kappa)$ and $\sin(2\kappa)$
in the elementary excitation spectrum $\Lambda_{\kappa}$ (\ref{1.07})
which arise from the three-site spin interactions,
may modify $\Lambda_{\kappa}$ drastically, 
leading to new ground-state phases.
We consider the case $J\ne 0$, $K\ne 0$, $D=E=0$;
for this case $\Lambda_{-\kappa}=\Lambda_{\kappa}$.
As long as $\vert K\vert<\vert J\vert /2$ the spinless fermion system may possess only two Fermi points,
whereas for sufficiently strong three-site interaction,
$\vert K\vert>\vert J\vert /2$,
the spinless fermion system may also possess four Fermi points.
The ground-state phase diagram obtained in Ref. \onlinecite{tj}
is shown in Fig.~\ref{fig01}.
Two different spin liquid phases reflect the importance of the existence of two versus four Fermi points,
as discussed already earlier \cite{tj}.
As a result,
in addition to the conventional quantum phase transition between the spin liquid I phase and the ferromagnetic phase,
the spin model may exhibit also quantum phase transitions
1) between the spin liquid I phase and the spin liquid II phase
and
2) between the spin liquid II phase and the ferromagnetic phase,
as well as a point where three phases meet
(see Fig.~\ref{fig01}).
The spin liquid I phase and the spin liquid II phase 
are characterized by a change in the power-law decay and oscillating factor 
of spin-spin correlations 
(see Eqs. (20) -- (23) in Ref.~\onlinecite{tj}).

To describe a zero-field quantum phase transition from the spin liquid I to the spin liquid II phase
at $\vert K\vert=2\vert J\vert$,
I.~Titvinidze and G.~I.~Japaridze \cite{tj} introduced the order parameter $\eta$
constructed from the average length $\overline{{\cal{L}}}$ of the ferromagnetic string in the ground state 
(see Eqs. (31) -- (33) in Ref. \onlinecite{tj}).
On the other hand,
P.~Lou {\it{et al.}} \cite{lou1} in their study of the (homogeneous) model (\ref{1.01})
with $J\ne 0$, $E\ne 0$, $D=K=0$
introduced the scalar chirality parameter
$O=-(1/N)\sum_{n}\langle s_{n-1}^xs_{n}^zs_{n+1}^y-s_{n-1}^ys_{n}^zs_{n+1}^x \rangle$
and calculated this quantity in the ground state at $\Omega=0$
(see Eqs. (10), (11) in Ref. \onlinecite{lou1}).
A nonzero value of $O$ signalizes the appearance of a new (chiral) spin liquid phase
which emerges when three-site interactions exceed a critical value.
Eliminating $XYZ-YZX$ terms from the Hamiltonian by the unitary transformation discussed above 
one arrives at the spin model
with $\tilde{J}=J/{\sqrt{2}}$, $\tilde{D}=-J/{\sqrt{2}}$, $\tilde{K}=E$, $\tilde{E}=0$
(i.e. the model similar to the one considered in Ref. \onlinecite{tj})
and $O\to\tilde{O}=-(1/N)\sum_{n}\langle s_{n-1}^xs_{n}^zs_{n+1}^x+s_{n-1}^ys_{n}^zs_{n+1}^y\rangle$.
The latter quantity $\tilde{O}$ apparently has no relation to the order parameter $\eta$ used in Ref.~\onlinecite{tj}.

It is also worth mentioning here 
that the spin model (\ref{1.01}) can be written 
as a one-dimensional model of hard-core bosons 
after introducing the on-site creation and annihilation operators
$s_n^{+}=s_n^x+{\rm{i}}s_n^y$ and $s_n^{-}=s_n^x-{\rm{i}}s_n^y$,
$s_n^z=s_n^{+}s_n^{-}-1/2$.
The hard-core boson model is obtained by taking the $U\to\infty$ limit of the boson Hubbard model.
With this mapping 
in the case of a conventional transverse $XX$ chain 
the ferromagnetic phase corresponds to the Mott insulator 
with $(1/N)\sum_n\langle s_n^{+}s_n^{-} \rangle =0$
or $(1/N)\sum_n\langle s_n^{+}s_n^{-} \rangle =1$
(the thermodynamic average taken at zero temperature, $T=0$),
whereas the spin liquid (spin liquid I) phase is the superfluid 
with $0< (1/N)\sum_n\langle s_n^{+}s_n^{-} \rangle <1$.
As a function of the field / chemical potential $\Omega$ 
the model displays two superfluid to Mott insulator transitions at $\Omega=\pm\vert J\vert$ \cite{sachdev}.
After switching on the three-site interaction $K\ne0$,
the picture remains qualitatively the same as long as $\vert K/J\vert<1/2$.
If $\vert K/J\vert$ exceeds 1/2 we face an additional transition 
which manifests itself as an extra cusp in the dependence 
of $(1/N)\sum_n\langle s_n^{+}s_n^{-} \rangle$ on $\Omega$.
Thus, following the ground-state average boson number per site 
(or $m_z=(1/N)\sum_n\langle s_n^z\rangle=(1/N)\sum_n\langle s_n^{+}s_n^{-} \rangle-1/2$)
as a function of $\Omega$ one may reproduce the various phases 
and the phase transitions between them shown in Fig. \ref{fig01}. 
An alternative way to follow the changes in the ground-state dependence of $m_z$ on $\Omega$ 
is to examine the ground-state susceptibility 
$\chi_{zz}=\partial m_z/\partial \Omega$ 
as a function of $\Omega$.
We notice that the ground-state dependence of $-\chi_{zz}$ on $\Omega$ 
is the same as that of $\rho(E=0)$ on $\Omega$,
where $\rho(E)=(1/N)\sum_{\kappa}\delta(E-\Lambda_{\kappa})$,
$\Lambda_{\kappa}=J\cos\kappa-(K/2)\cos(2\kappa)+\Omega$ 
is the one-particle density of states.
As a result, 
the ground-state dependence $\chi_{zz}$ vs $\Omega$ exhibits a square-root van Hove singularity 
along the lines separating different phases in Fig. \ref{fig01}.
(The only exception are the two points $K=\pm\vert J\vert/2$, $\Omega=\mp\vert J\vert +K/2$
at which $\Lambda_{\kappa}\propto \kappa^4$
and therefore $\chi_{zz}$ displays a van Hove singularity with the exponent 3/4.)
The divergence of the uniform static $zz$ susceptibility 
implies a ``ferromagnetic'' character of the associated phase transitions.

To summarize,
there is no doubt that while ``the two Fermi points spinless fermions''
transform into ``the four Fermi points spinless fermions''
some noticeable changes in the properties of the spin model should take place,
however, a transparent quantity associated with this modification of the Fermi surface topology
which may play the role of the order parameter is still lacking.

Finally,
it is worth noting the studies
on the one-dimensional Hubbard model with next-nearest-neighbor hopping
because the noninteracting limit of that model resembles Eq. (\ref{1.07})
(see, e.g., Ref. \onlinecite{hubbard} and references therein).
In contrast to those studies,
we calculate two- and many-particle correlation functions
(although for a system of noninteracting spinless fermions (\ref{1.07}))
which are related to two-spin correlation functions
for an interacting quantum spin system.

In our study of dynamic properties of the spin model we focus on the dynamic structure factor
\begin{eqnarray}
\label{1.08}
S_{AB}(\kappa,\omega)=\sum_{l=1}^N\exp\left(-{\rm{i}}\kappa l\right)
\int_{-\infty}^{\infty}{\rm{d}}t\exp\left({\rm{i}}\omega t\right)
\left\langle
\left(A_n(t)-\langle A \rangle\right)
\left(B_{n+l}(0)-\langle B \rangle\right)
\right\rangle,
\end{eqnarray}
where $A_n$, $B_n$ are some local operators attached to the site $n$
(like $s_n^{\alpha}$, $\alpha=x,y,z$ or $d_n^{(1)}=s_n^xs_{n+1}^x+s_n^ys_{n+1}^y$ etc.),
$A_{n}(t)=\exp({\rm{i}}Ht)A_{n}\exp(-{\rm{i}}Ht)$,
$\langle (\ldots)\rangle
={\rm{Tr}}(\exp(-\beta H)(\ldots))/{\rm{Tr}}\exp(-\beta H)$,
$\langle A\rangle=(1/N)\sum_n\langle A_n\rangle$.
Knowing the dynamic structure factors we can find the corresponding dynamic susceptibilities
according to well known relations
(see, e.g., Ref. \onlinecite{zubarev}).

In the next section
(Sec.~\ref{sec2})
we report a closed-form expression for the $zz$ dynamic structure factor $S_{zz}(\kappa,\omega)$
(i.e. $A_{n}=B_{n}=s_n^z$ in (\ref{1.08}))
and for some similar dynamic structure factors $S_{I}(\kappa,\omega)$
(see Eqs. (\ref{2.03}), (\ref{2.04}) below)
all of which are governed by two-fermion (particle-hole) excitations.
Then,
in Sec.~\ref{sec3},
we discuss  general properties of the two-fermion excitation continuum
focusing on spectral boundaries, soft modes, singularities etc.
We also contrast generic and specific features of various two-fermion dynamic quantities.
In Sec.~\ref{sec4} we examine many-fermion dynamic quantities
focusing, in particular, on the $xx$ dynamic structure factor $S_{xx}(\kappa,\omega)$
(i.e. $A_{n}=B_{n}=s_n^x$ in (\ref{1.08})).
We report exact analytical results
1) in the high-temperature regime $\beta=0$ ($T\to\infty$)
and
2) in the strong-field regime in the ground state
as well as precise numerical results for arbitrary temperatures.
Finally, in Sec.~\ref{sec5}, we summarize our findings.
Some selected results of the present study were announced in a conference paper \cite{kdsv}.

\section{$zz$ dynamic structure factor and some other two-fermion dynamic structure factors}
\label{sec2}
\setcounter{equation}{0}

We start with the calculation of the transverse dynamic structure factor $S_{zz}(\kappa,\omega)$
that corresponds to $A_n=B_n=s_n^z$ in Eq. (\ref{1.08}).
Since according to Eq. (\ref{1.04}) $s_n^z=c_n^{\dagger}c_n-1/2$
the calculation of the transverse dynamic structure factor is very simple \cite{niemejer,taylor}.
Using Eqs. (\ref{1.06}), (\ref{1.07}) and the Wick-Bloch-de Dominicis theorem
we end up with the result
\begin{eqnarray}
\label{2.01}
S_{zz}(\kappa,\omega)
=\int_{-\pi}^{\pi}{\rm{d}}\kappa_1
n_{\kappa_1}\left(1-n_{\kappa_1+\kappa}\right)
\delta\left(\omega+\Lambda_{\kappa_1}-\Lambda_{\kappa_1+\kappa}\right)
\nonumber\\
=\left.\sum_{\kappa_1^{\star}}\frac{n_{\kappa_1}\left(1-n_{\kappa_1+\kappa}\right)}
{\left\vert\frac{\partial}{\partial\kappa_1}\left(\Lambda_{\kappa_1}-\Lambda_{\kappa_1+\kappa}\right)\right\vert}
\right\vert_{\kappa_1=\kappa_1^{\star}}.
\end{eqnarray}
Here
$n_{\kappa}=1/\left(1+\exp(\beta\Lambda_{\kappa})\right)$
is the Fermi function,
and $\{\kappa_1^{\star}\}$ are the solutions of the equation
\begin{eqnarray}
\label{2.02}
\omega+\Lambda_{\kappa_1^{\star}}-\Lambda_{\kappa_1^{\star}+\kappa}=0.
\end{eqnarray}

There are more local spin operators which in fermionic representation are given by a product of two Fermi operators,
\begin{eqnarray}
\label{2.03}
d^{(1)}_n=s_n^xs_{n+1}^x+s_n^ys_{n+1}^y
=\frac{1}{2}\left(c_n^{\dagger}c_{n+1}-c_nc^{\dagger}_{n+1}\right),
\nonumber\\
d^{(2)}_n=s_n^xs_{n+1}^y-s_n^ys_{n+1}^x
=\frac{{\rm{i}}}{2}\left(c_n^{\dagger}c_{n+1}+c_nc^{\dagger}_{n+1}\right),
\nonumber\\
t^{(1)}_n=s_n^xs_{n+1}^zs_{n+2}^x+s_n^ys_{n+1}^zs_{n+2}^y
=-\frac{1}{4}\left(c_n^{\dagger}c_{n+2}-c_nc^{\dagger}_{n+2}\right),
\nonumber\\
t^{(2)}_n=s_n^xs_{n+1}^zs_{n+2}^y-s_n^ys_{n+1}^zs_{n+2}^x
=-\frac{{\rm{i}}}{4}\left(c_n^{\dagger}c_{n+2}+c_nc^{\dagger}_{n+2}\right)
\end{eqnarray}
etc.
The correlation functions
$\langle d^{(1)}_n(t) d^{(1)}_{n+l}(0)\rangle$,
$\langle d^{(2)}_n(t) d^{(2)}_{n+l}(0)\rangle$,
$\langle t^{(1)}_n(t) t^{(1)}_{n+l}(0)\rangle$,
$\langle t^{(2)}_n(t) t^{(2)}_{n+l}(0)\rangle$
and the corresponding structure factors
$S_{J}(\kappa,\omega)$,
$S_{D}(\kappa,\omega)$,
$S_{K}(\kappa,\omega)$,
$S_{E}(\kappa,\omega)$
therefore can be again easily calculated with the result
\begin{eqnarray}
\label{2.04}
S_{I}(\kappa,\omega)
=\int_{-\pi}^{\pi}{\rm{d}}\kappa_1
B_{I}(\kappa_1,\kappa_1+\kappa)
C(\kappa_1,\kappa_1+\kappa)
\delta\left(\omega-E(\kappa_1,\kappa_1+\kappa)\right)
\nonumber\\
=\left.\sum_{\kappa_1^{\star}}\frac{B_{I}(\kappa_1,\kappa_1+\kappa)C(\kappa_1,\kappa_1+\kappa)}
{\left\vert\frac{\partial}{\partial\kappa_1}E(\kappa_1,\kappa_1+\kappa)\right\vert}
\right\vert_{\kappa_1=\kappa_1^{\star}},
\nonumber\\
B_{J}(\kappa_1,\kappa_2)
=\cos^2\frac{\kappa_1+\kappa_2}{2},
\nonumber\\
B_{D}(\kappa_1,\kappa_2)
=\sin^2\frac{\kappa_1+\kappa_2}{2},
\nonumber\\
B_{K}(\kappa_1,\kappa_2)
=\frac{1}{4}\cos^2\left(\kappa_1+\kappa_2\right),
\nonumber\\
B_{E}(\kappa_1,\kappa_2)
=\frac{1}{4}\sin^2\left(\kappa_1+\kappa_2\right),
\nonumber\\
C(\kappa_1,\kappa_2)
=n_{\kappa_1}\left(1-n_{\kappa_2}\right),
\nonumber\\
E(\kappa_1,\kappa_2)
=-\Lambda_{\kappa_1}+\Lambda_{\kappa_2}.
\end{eqnarray}
We note that $S_{zz}(\kappa,\omega)$ (\ref{2.01}) is also given by Eq. (\ref{2.04})
with $B_{zz}(\kappa_1,\kappa_2)=1$.

\begin{figure}
\epsfig{file = 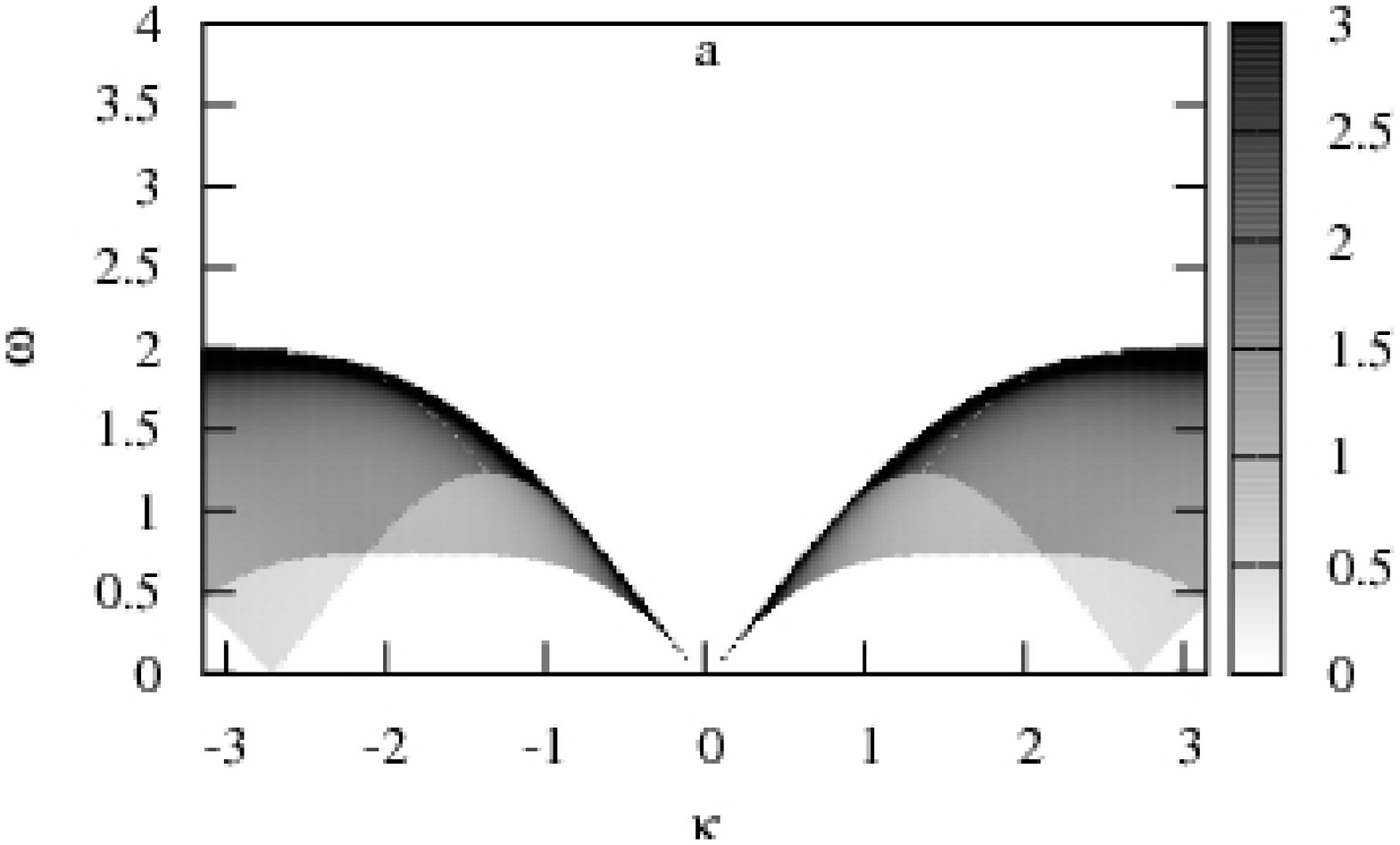, width = 0.32\linewidth}
\epsfig{file = 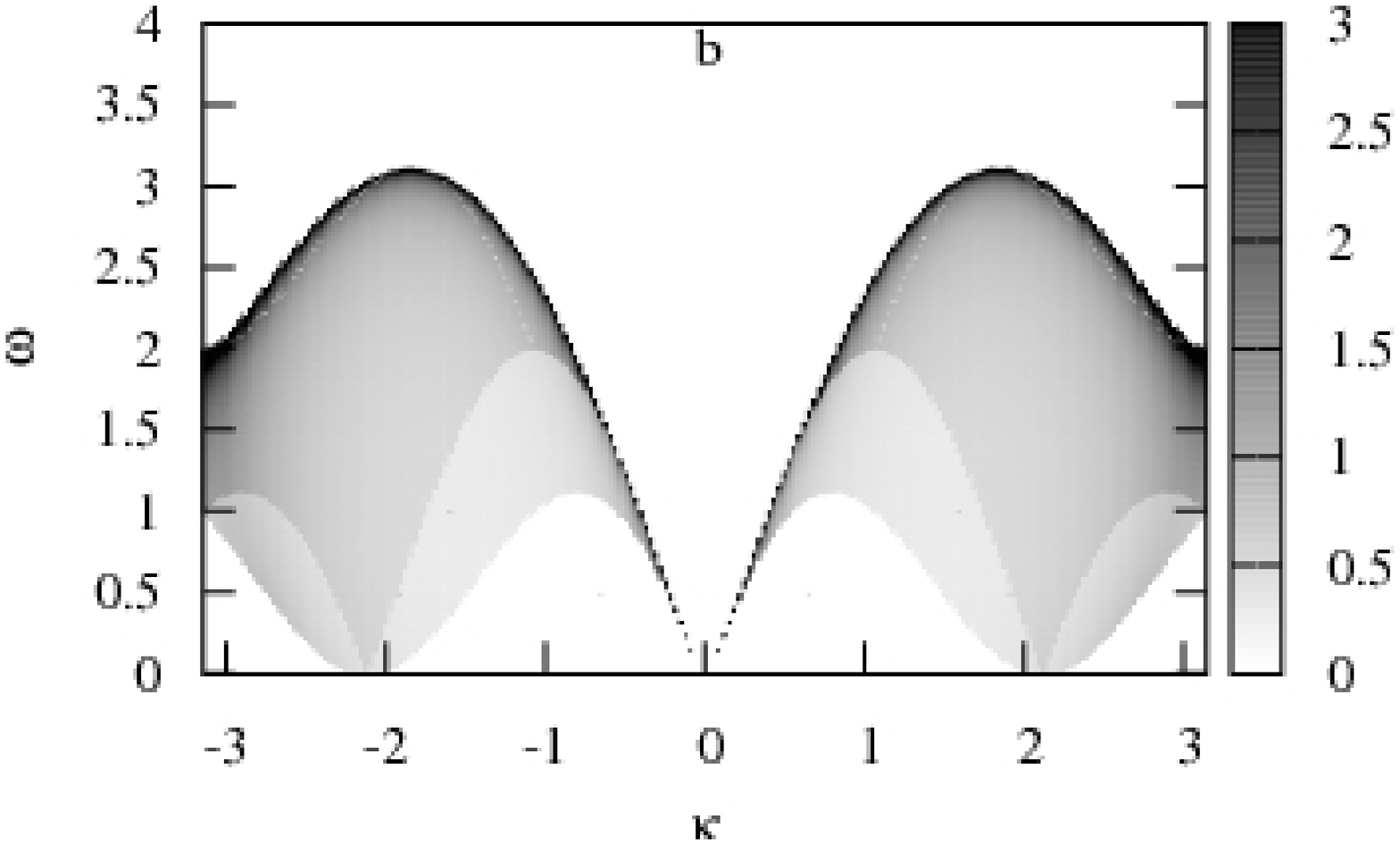, width = 0.32\linewidth}
\epsfig{file = 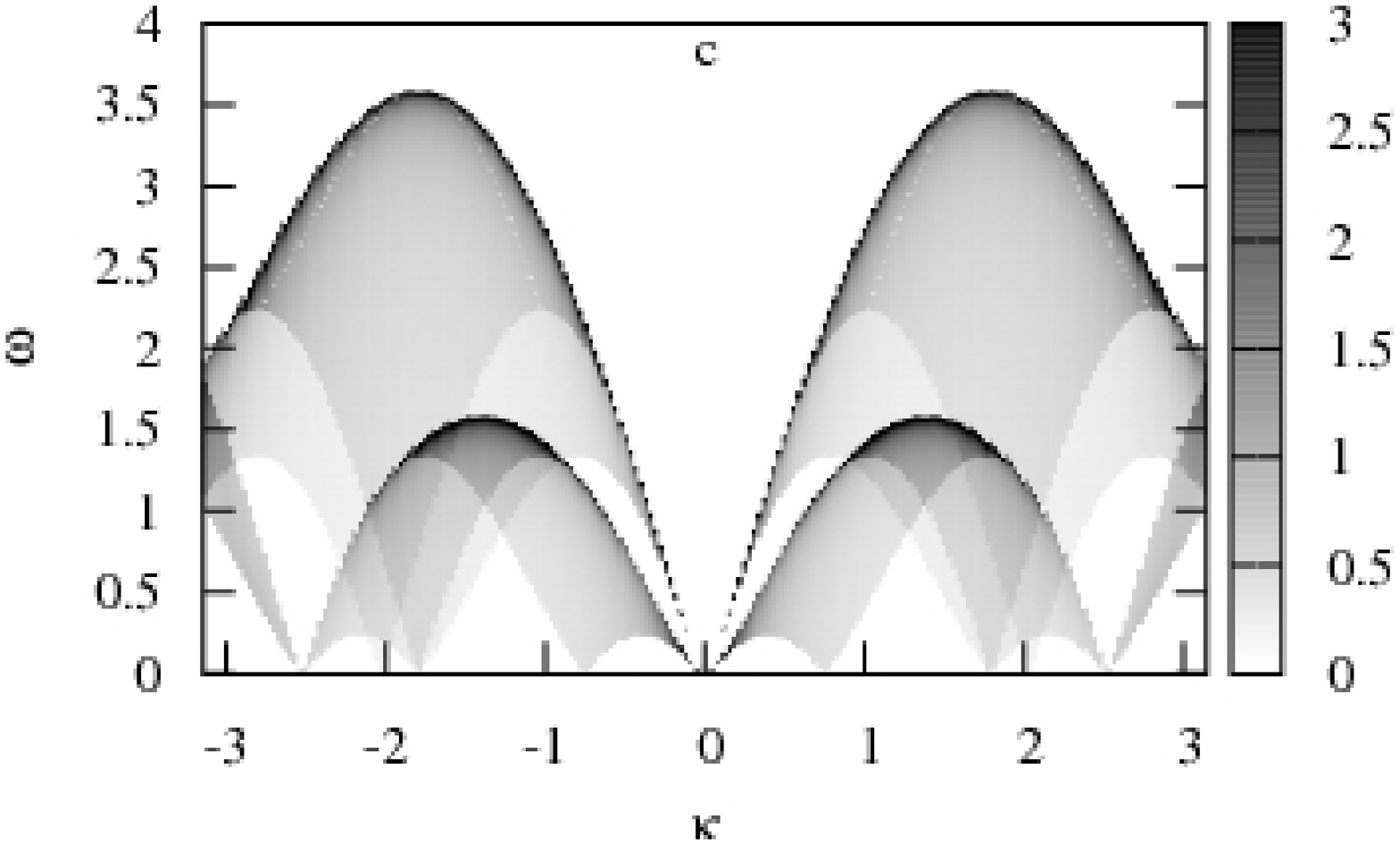, width = 0.32\linewidth}\\
\epsfig{file = 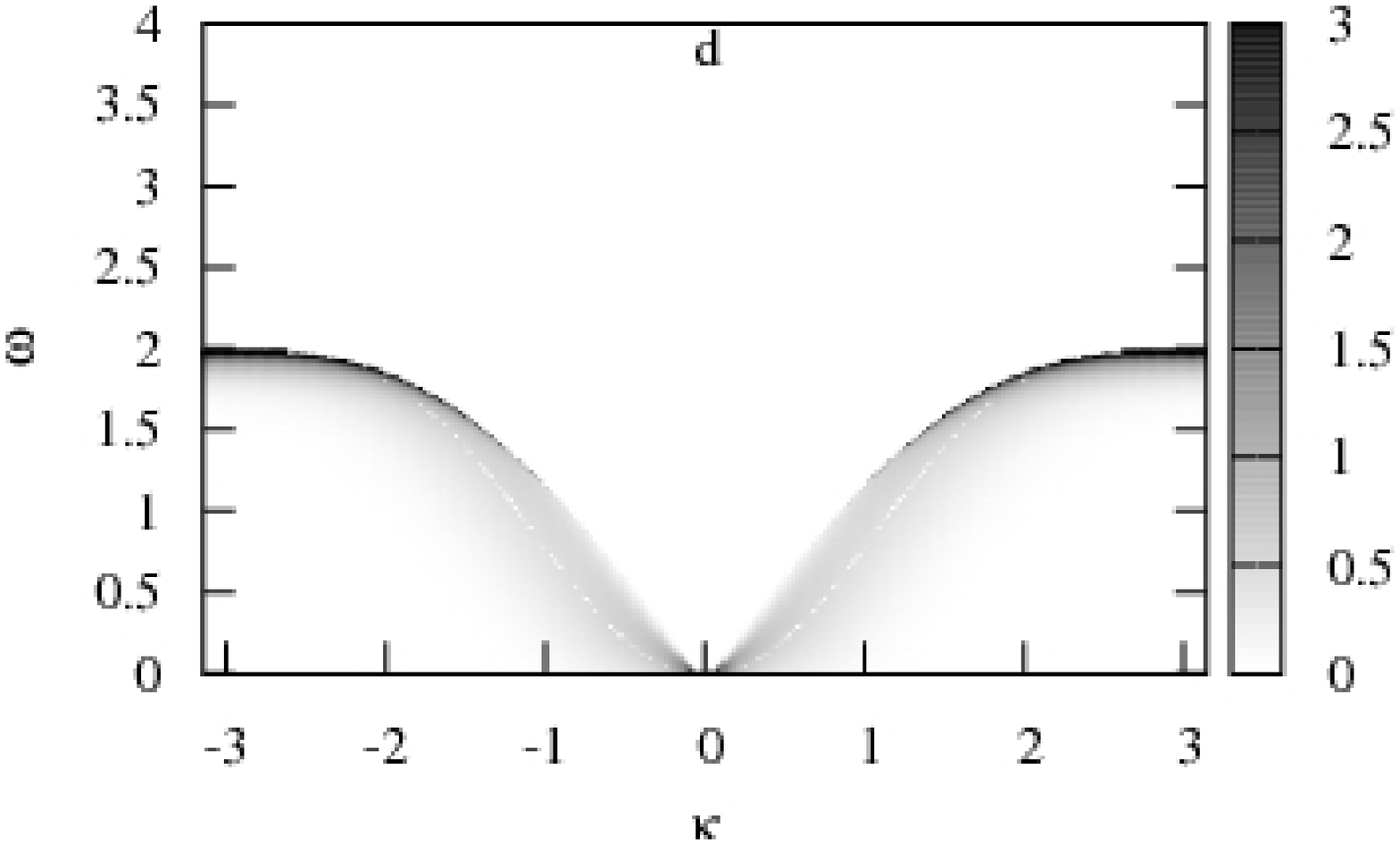, width = 0.32\linewidth}
\epsfig{file = 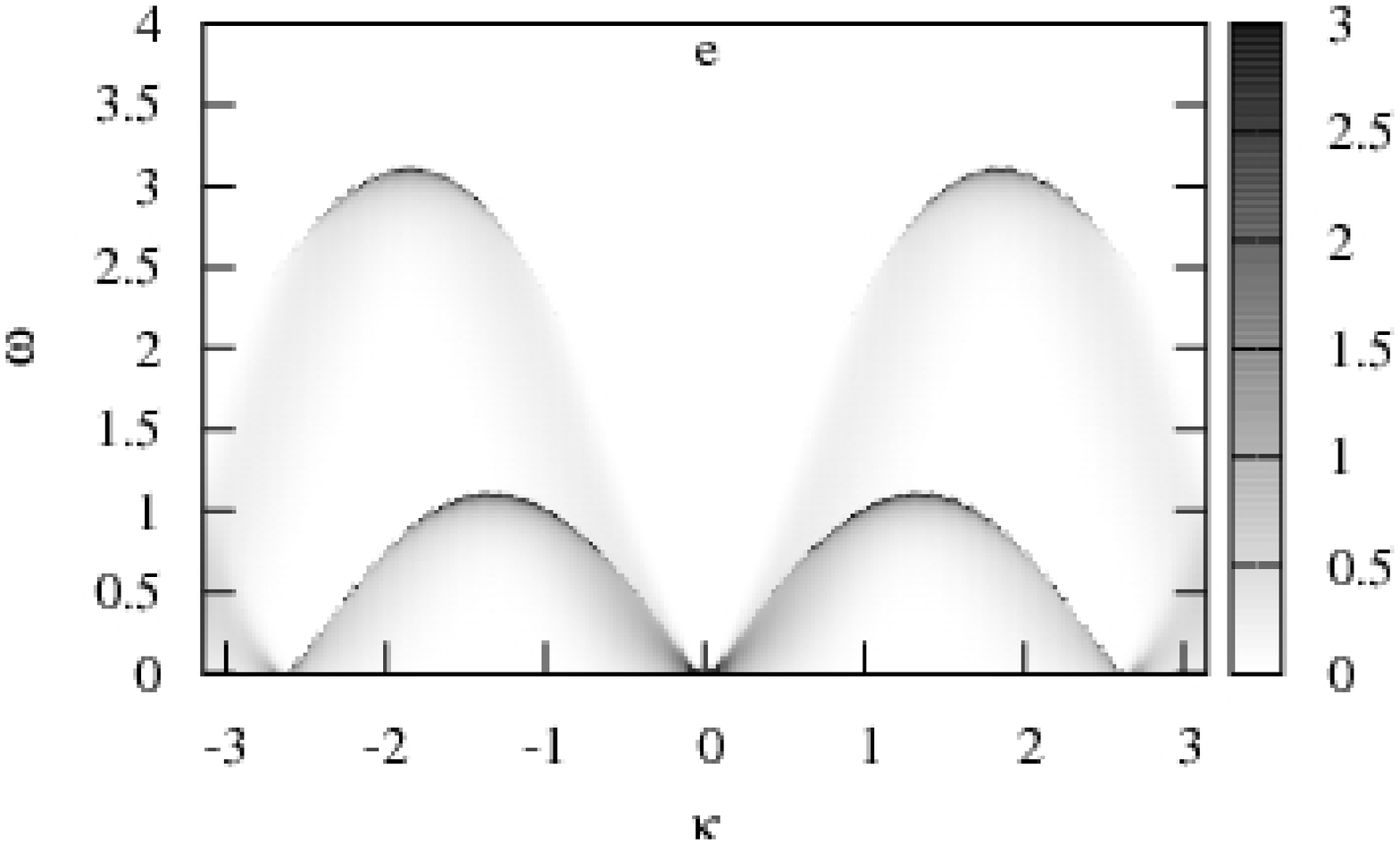, width = 0.32\linewidth}
\epsfig{file = 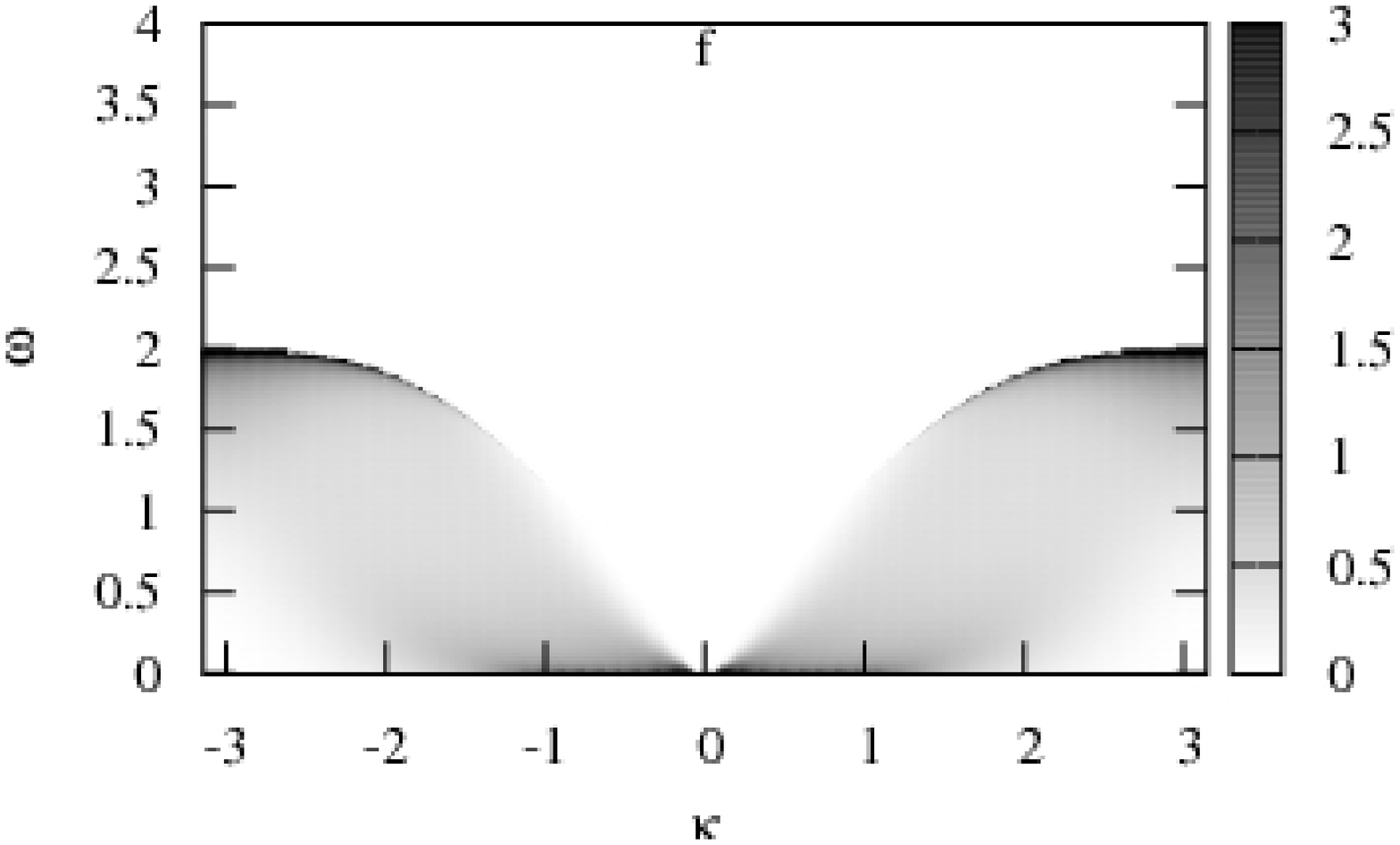, width = 0.32\linewidth}
\caption{$S_{zz}(\kappa,\omega)$
for the model (\ref{1.01})
with $J=1$, $D=E=0$,
$K=0.5$, $\Omega=0$ (panel a),
$K=2$, $\Omega=0$ (panel b),
$K=2.5$, $\Omega=0$ (panel c),
$K=0.5$, $\Omega=1.25$ (panel d),
$K=2$, $\Omega=-1.125$ (panel e),
$K=0.5$, $\Omega=-0.75$ (panel f).
Panels a, b, c refer to the ground state ($T=0$),
panels d, e, f refer to the low temperature $\beta=10$.}
\label{fig02}
\end{figure}

\begin{figure}
\epsfig{file = 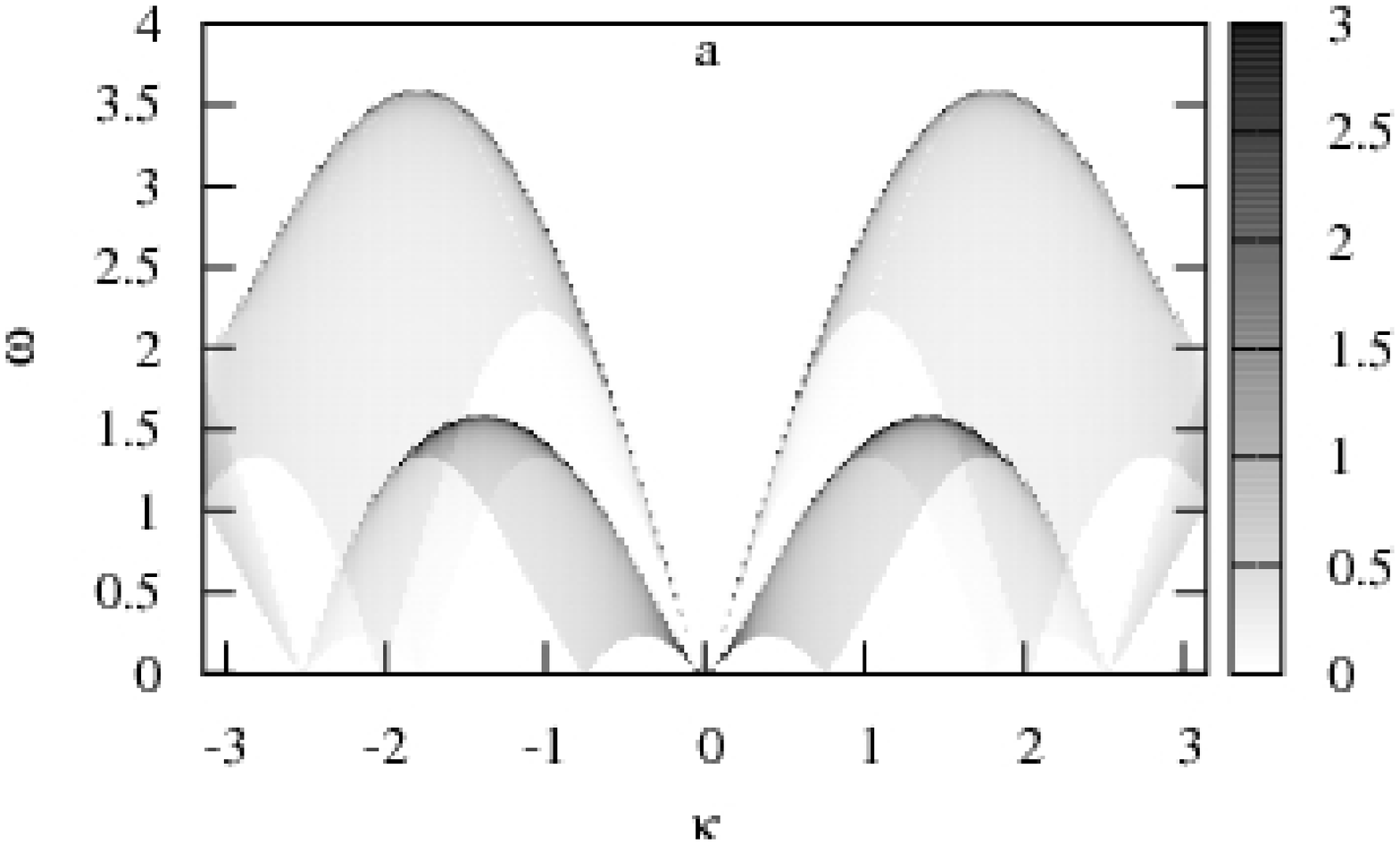, width = 0.32\linewidth}\\
\epsfig{file = 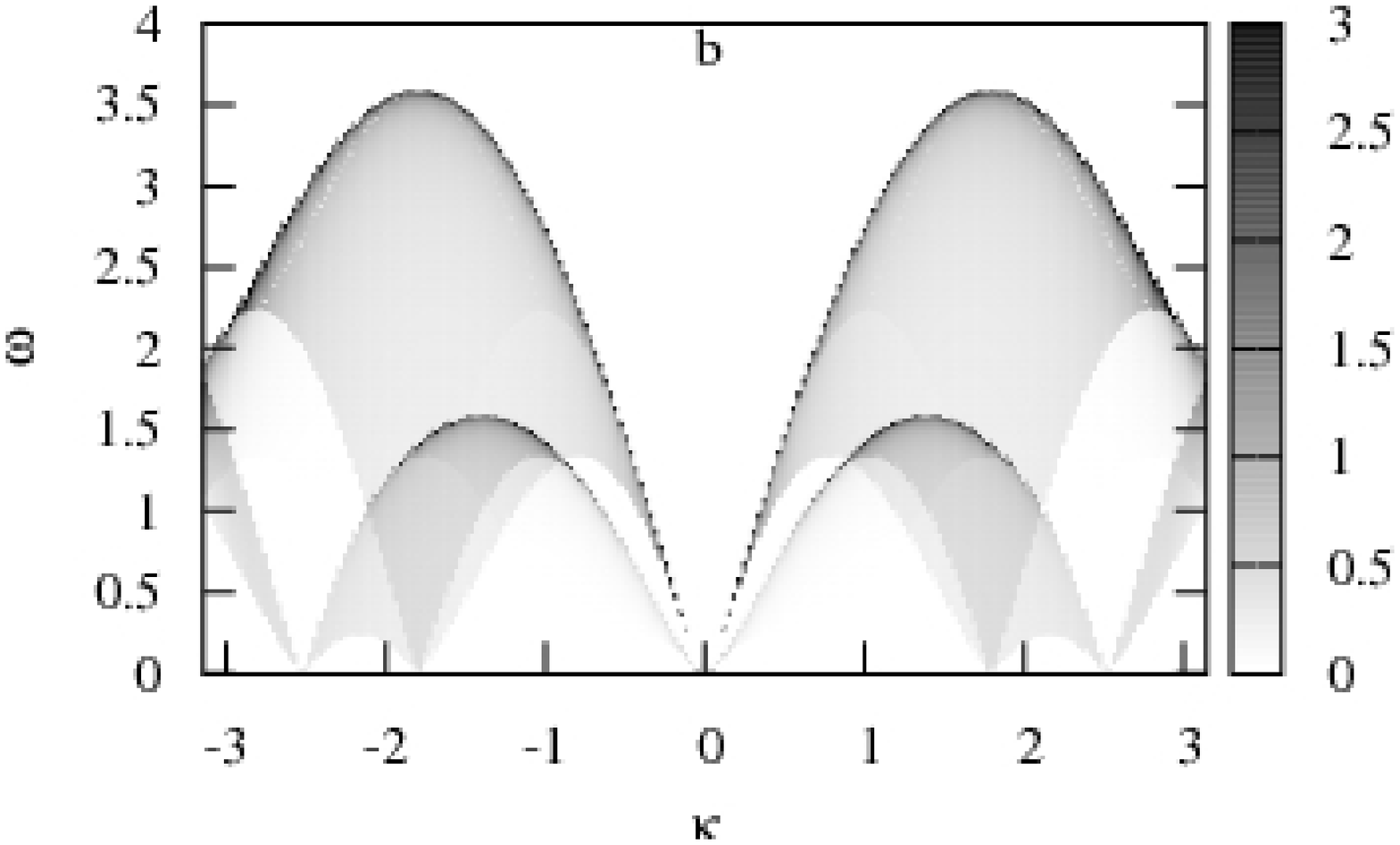, width = 0.32\linewidth}\\
\epsfig{file = 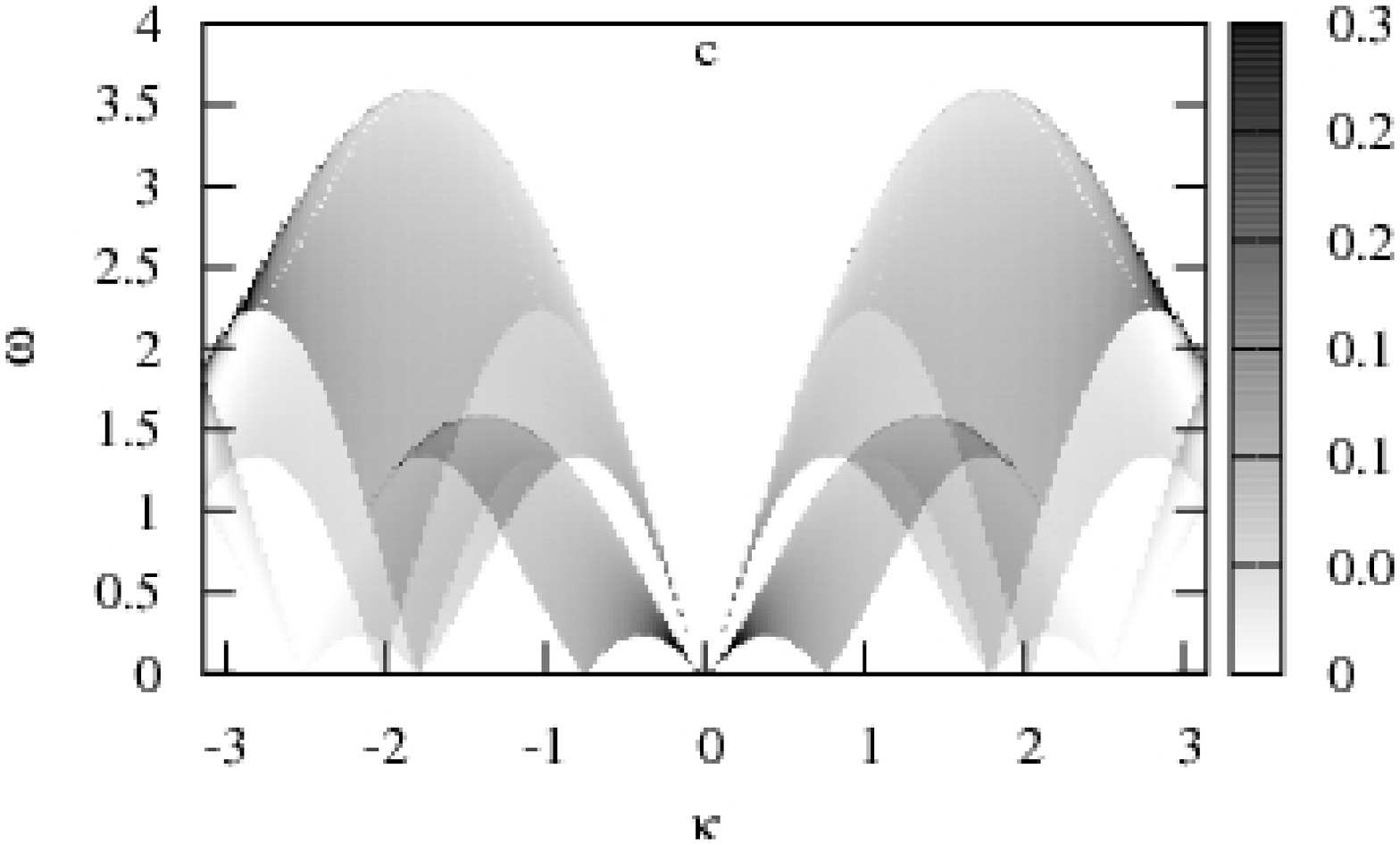, width = 0.32\linewidth}\\
\epsfig{file = 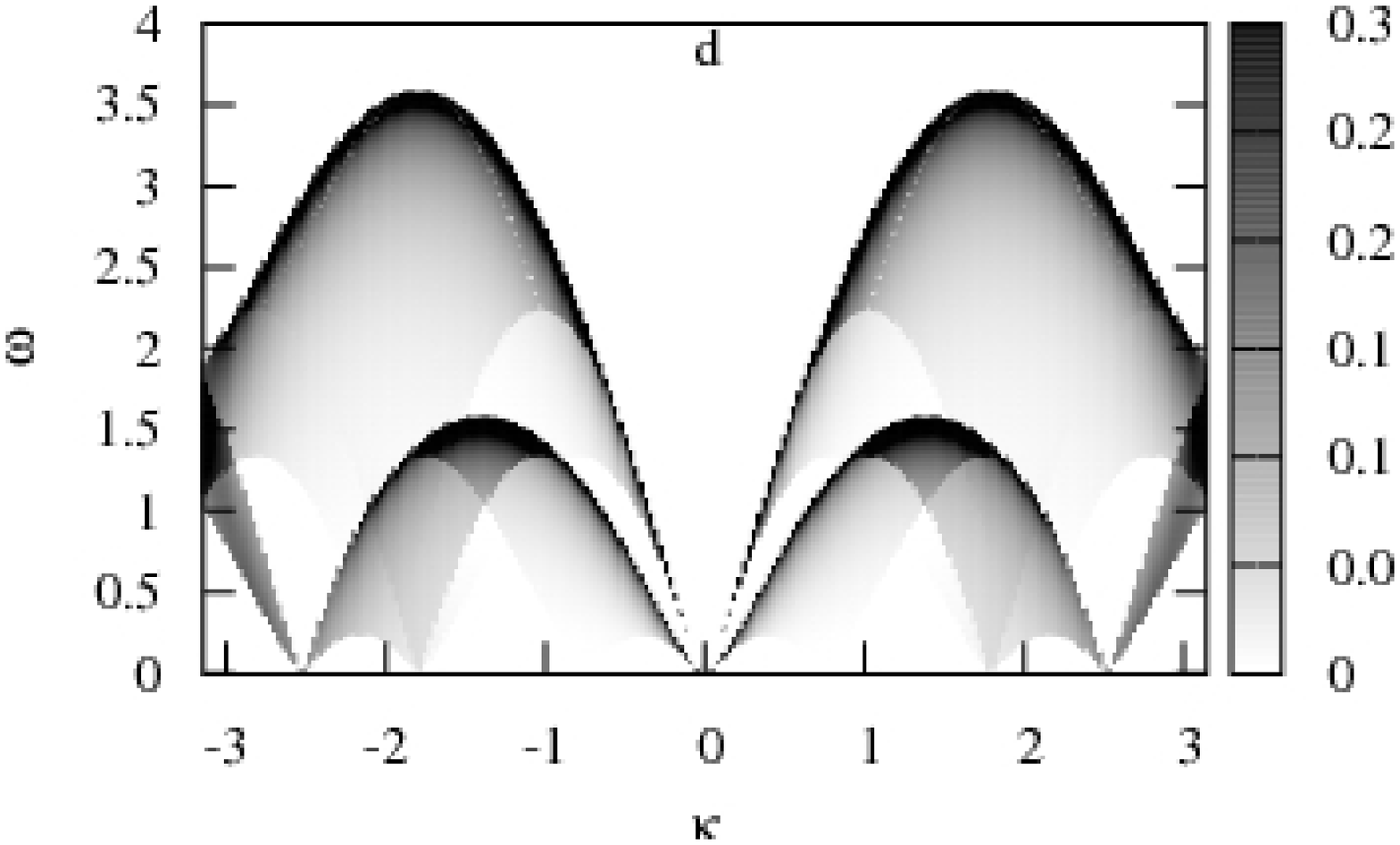, width = 0.32\linewidth}
\caption{Two-fermion dynamic structure factors
$S_{J}(\kappa,\omega)$ (panel a),
$S_{D}(\kappa,\omega)$ (panel b),
$S_{K}(\kappa,\omega)$ (panel c),
$S_{E}(\kappa,\omega)$ (panel d)
for the model (\ref{1.01})
with $J=1$, $D=E=0$, $K=2.5$, $\Omega=0$
at $T=0$.}
\label{fig03}
\end{figure}

\begin{figure}
\epsfig{file = 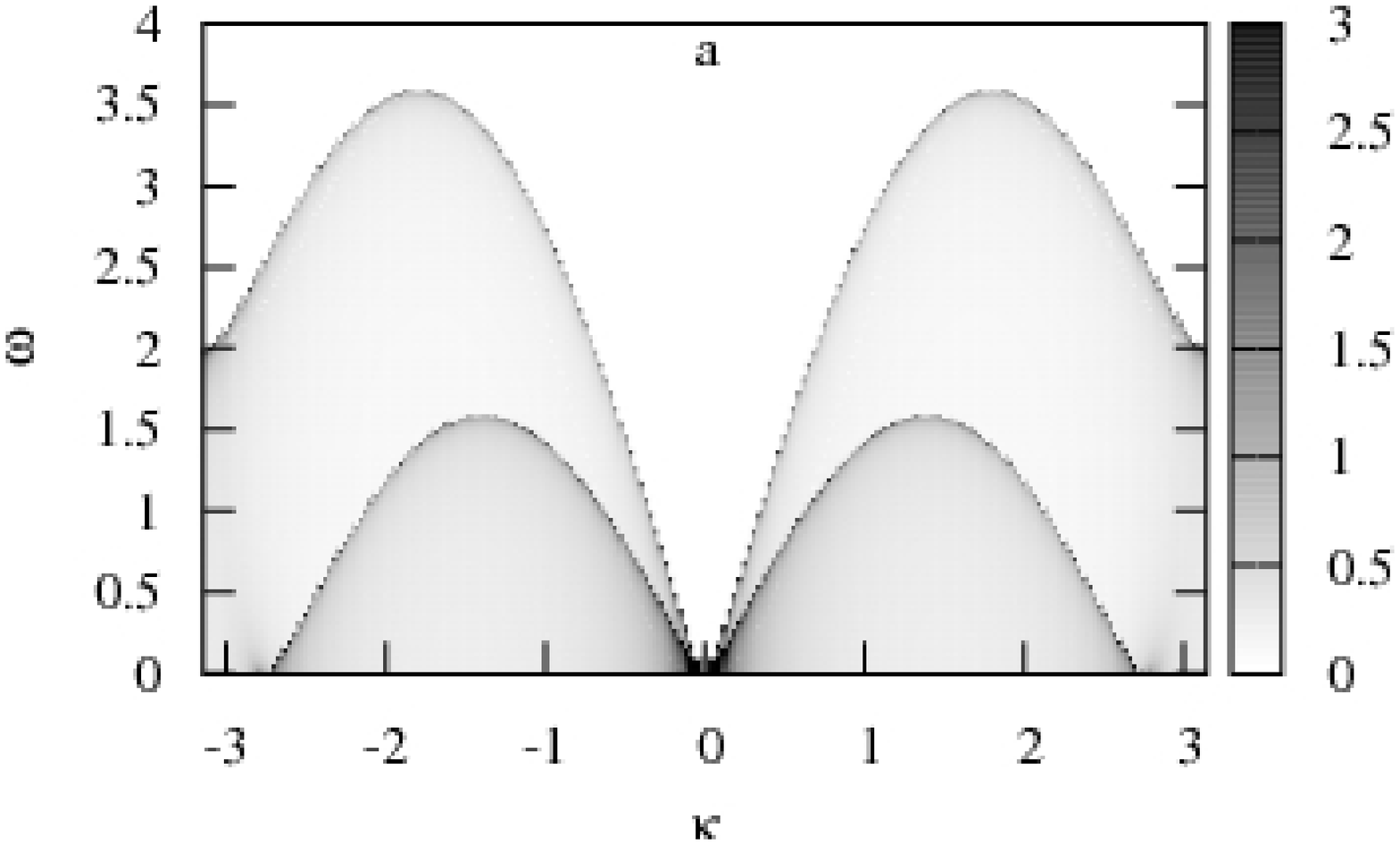, width = 0.32\linewidth}
\epsfig{file = 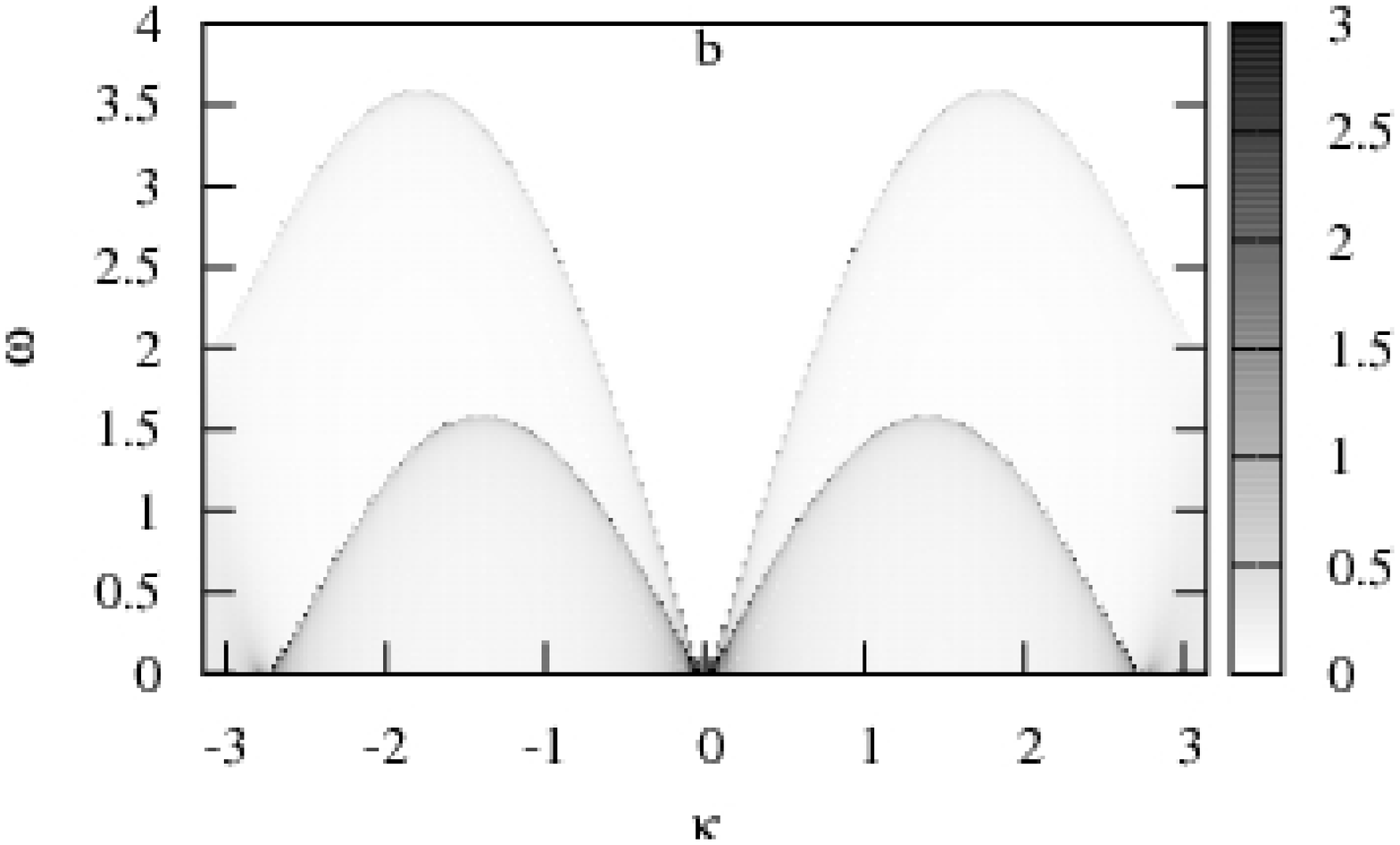, width = 0.32\linewidth}
\epsfig{file = 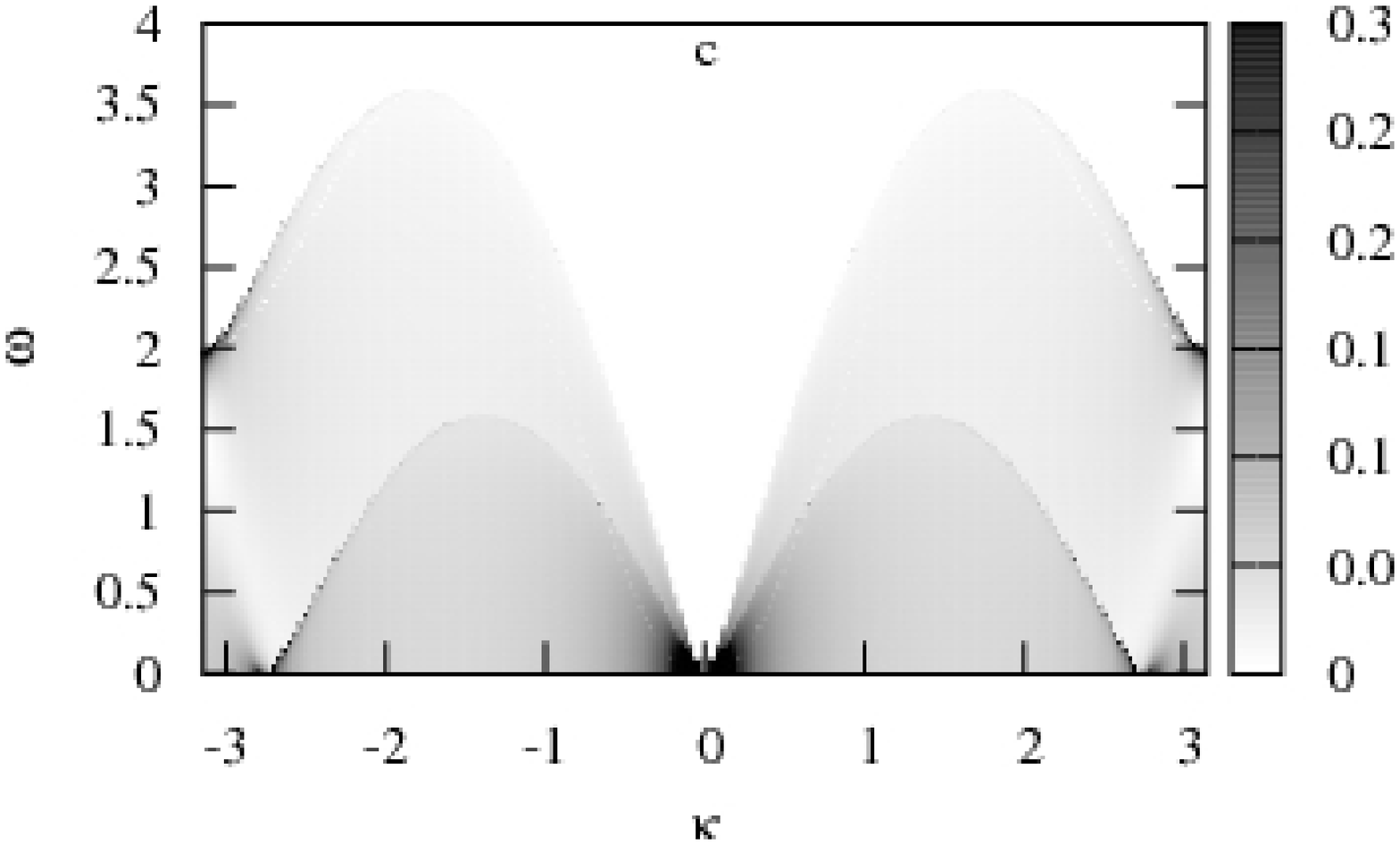, width = 0.32\linewidth}\\
\caption{Two-fermion dynamic structure factors
$S_{zz}(\kappa,\omega)$ (panel a),
$S_{J}(\kappa,\omega)$ (panel b),
$S_{K}(\kappa,\omega)$ (panel c)
for the model (\ref{1.01})
with $J=1$, $D=E=0$, $K=2.5$
at $T\to \infty$.
Note that for infinite temperature 
the two-fermion dynamic structure factors are field-independent.}
\label{fig04}
\end{figure}

In Figs. \ref{fig02}, \ref{fig03}, \ref{fig04}
we show the gray-scale plots 
for the different two-fermion dynamic structure factors (\ref{2.01}), (\ref{2.04})
for several representative sets of the Hamiltonian parameters,
at zero and infinite temperatures.
Comparing the closed-form expressions (\ref{2.01}), (\ref{2.04})
and the gray-scale plots in Figs. \ref{fig02}, \ref{fig03}, \ref{fig04}
we conclude that
1) the generic properties of all these two-fermion dynamic quantities
are controlled by the $\delta$-function containing $E(\kappa_1,\kappa_2)$ 
in Eqs. (\ref{2.01}), (\ref{2.04}) in the high-temperature limit $T\to\infty$ ($\beta\to 0$),
whereas in the low-temperature limit $T\to 0$ ($\beta\to\infty$) 
the factor containing Fermi functions,
$C(\kappa_1,\kappa_2)$,
becomes important in addition
and
2) the specific properties of the various two-fermion dynamic quantities
arise only owing to different functions $B_{I}(\kappa_1,\kappa_2)$.
In the next section we further examine the two-fermion dynamic structure factors,
revealing their similarities and contrasting their differences.

\section{Two-fermion excitation continuum: Generic versus specific properties}
\label{sec3}
\setcounter{equation}{0}

Let us discuss the properties of the two-fermion excitation continuum
which is probed by a number of dynamic quantities like
the transverse dynamic structure factor $S_{zz}(\kappa,\omega)$,
the dimer dynamic structure factor $S_{J}(\kappa,\omega)$ etc.
For the model under consideration here,
(\ref{1.01}) with coupling constants $D=E=0$,
the elementary excitation spectrum (\ref{1.07}) 
differs from that of the standard homogeneous $XX$ chain 
by the $\cos(2\kappa)$-term.
That term has important consequences 
which we explore by generalizing the work of J.~H.~Taylor and G.~M\"{u}ller \cite{taylor} 
on conventional $XY$ chains.

We start with the high-temperature limit $T\to\infty$.
A two-fermion dynamic quantity (\ref{2.04}) may have a nonzero value 
at a point $(\kappa,\omega)$ 
in the wavevector -- frequency plane
(we assume $\omega\ge 0$, $-\pi\le \kappa <\pi$)
if
\begin{eqnarray}
\omega=E(\kappa_1,\kappa_2)=-\Lambda_{\kappa_1}+\Lambda_{\kappa_2},
\;\;\;
\kappa=-\kappa_1+\kappa_2 ({\rm{mod}}(2\pi)),
\nonumber\\
\Lambda_\kappa=J\cos\kappa-\frac{K}{2}\cos(2\kappa)+\Omega,
\label{3.01}
\end{eqnarray}
where $-\pi\le\kappa_1 <\pi$.
We rewrite the function $E(\kappa_1,\kappa_1+\kappa)$ in the form
\begin{eqnarray}
E(\kappa_1,\kappa_1+\kappa)
=2\sin\frac{\kappa}{2}\sin\left(\frac{\kappa}{2}+\kappa_1\right)
\left(-J+2K\cos\frac{\kappa}{2}\cos\left(\frac{\kappa}{2}+\kappa_1\right)\right)
\label{3.02}
\end{eqnarray}
and solve the equation
$\partial E(\kappa_1,\kappa_1+\kappa)/\partial\kappa_1 =0$
with respect to $\kappa_1$,
or more precisely with respect to $x=\cos\left(\kappa/2+\kappa_1\right)$,
to find
\begin{eqnarray}
x^{\pm}=\frac{J}{8K\cos\frac{\kappa}{2}}
\pm\sqrt{\left(\frac{J}{8K\cos\frac{\kappa}{2}}\right)^2+\frac{1}{2}}.
\label{3.03}
\end{eqnarray}
For $\vert K/J\vert<1/2$ there is only one pair of $\kappa_1$ 
which solves (\ref{3.03}) and fulfills the condition $\vert x\vert\le 1$,
thus yielding a stationary point of the function $E(\kappa_1,\kappa_1+\kappa)$.
That pair is given by
$\tilde{\kappa}^{-}_1=\arccos x^--\kappa/2$, 
$\tilde{\kappa}^{-}_1=-\arccos x^--\kappa/2$
for $JK>0$,
and by 
$\tilde{\kappa}^{+}_1=\arccos x^+-\kappa/2$, 
$\tilde{\kappa}^{+}_1=-\arccos x^+-\kappa/2$
for $JK<0$.
In the opposite case  $\vert K/J\vert>1/2$ there are two such pairs,
$\tilde{\kappa}^{-}_1=\arccos x^--\kappa/2$, 
$\tilde{\kappa}^{-}_1=-\arccos x^--\kappa/2$
and
$\tilde{\kappa}^{+}_1=\arccos x^+-\kappa/2$, 
$\tilde{\kappa}^{+}_1=-\arccos x^+-\kappa/2$.
As a result,
the upper boundary $\omega_{u}(\kappa)$ of the two-fermion excitation continuum
is given by
\begin{eqnarray}
\Omega_{-}(\kappa)
=
\vert E(\tilde{\kappa}_1^-,\tilde{\kappa}_1^-+\kappa)\vert, \;\;\; JK > 0
\label{3.04}
\end{eqnarray}
or
\begin{eqnarray}
\Omega_{+}(\kappa)
=
\vert E(\tilde{\kappa}_1^+,\tilde{\kappa}_1^++\kappa)\vert, \;\;\; JK < 0.
\label{3.05}
\end{eqnarray}
Note that for $\vert K/J\vert>1/2$
the first derivative of $E(\kappa_1,\kappa_1+\kappa)$ with respect to $\kappa_1$ is zero also
along $\Omega_{+}(\kappa)<\Omega_{-}(\kappa)=\omega_{u}(\kappa)$ for $JK > 0$
or
along $\Omega_{-}(\kappa)<\Omega_{+}(\kappa)=\omega_{u}(\kappa)$ for $JK < 0$.

\begin{figure}
\epsfig{file = 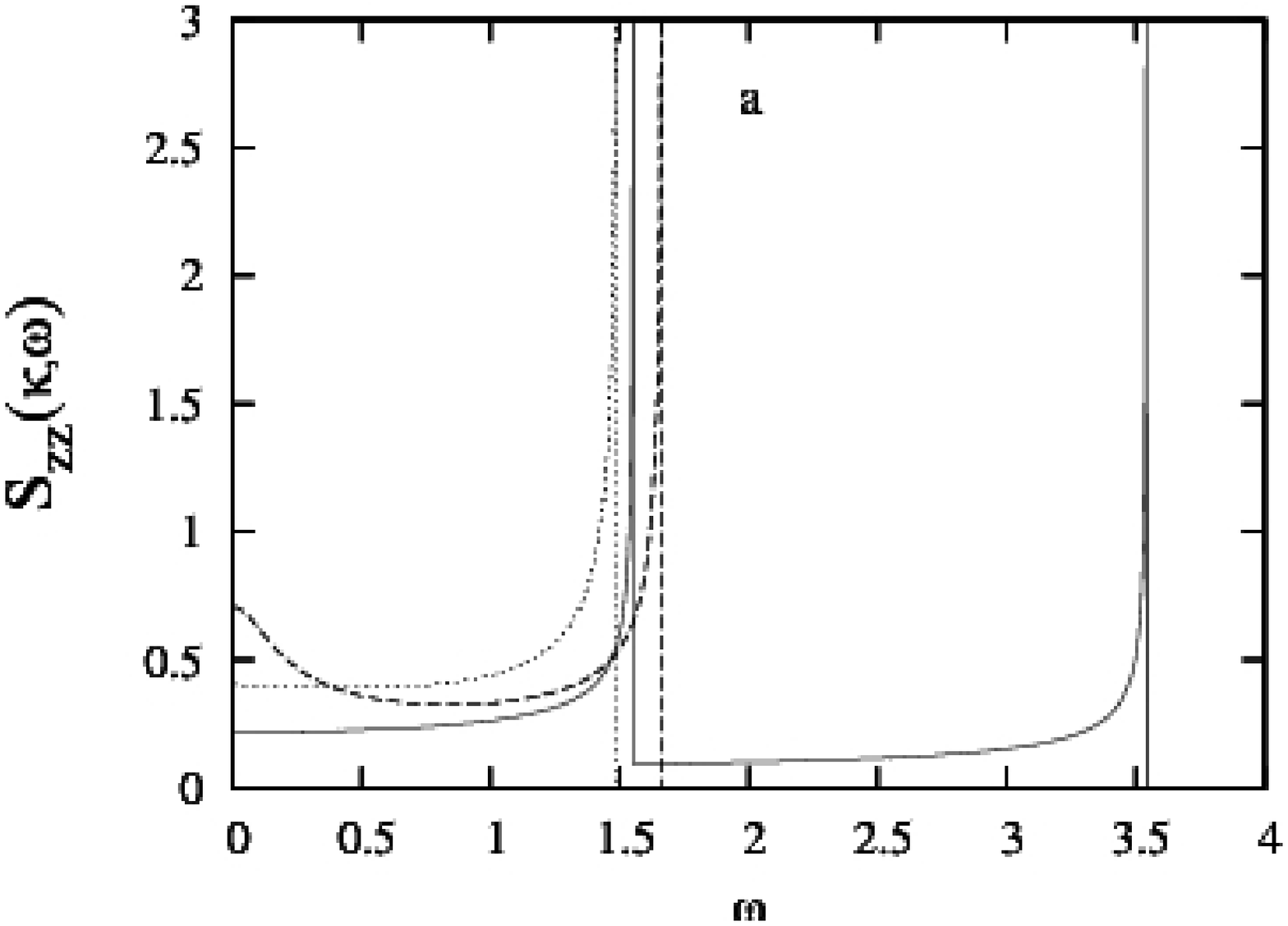, width = 0.32\linewidth}
\epsfig{file = 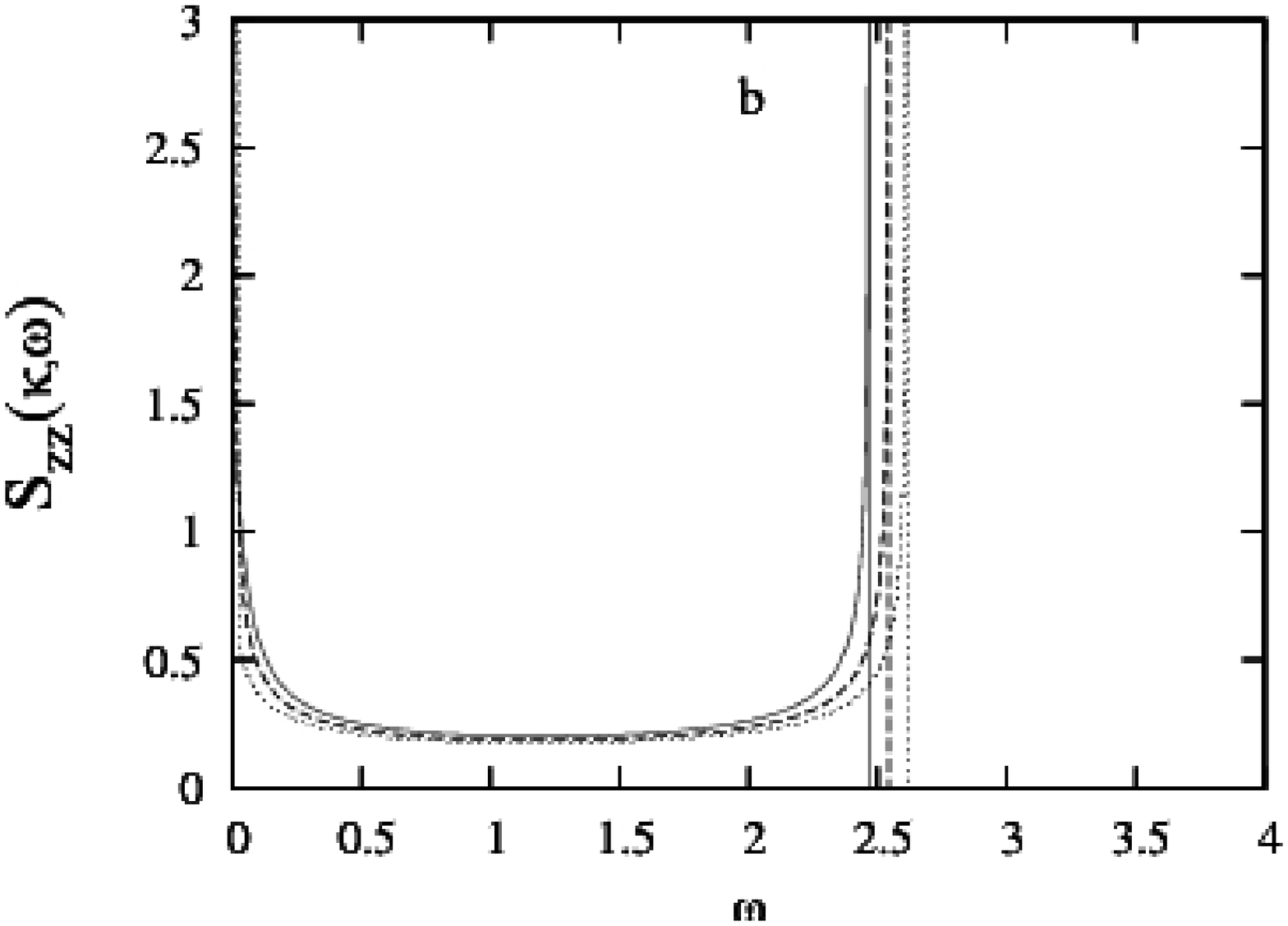, width = 0.32\linewidth}
\epsfig{file = 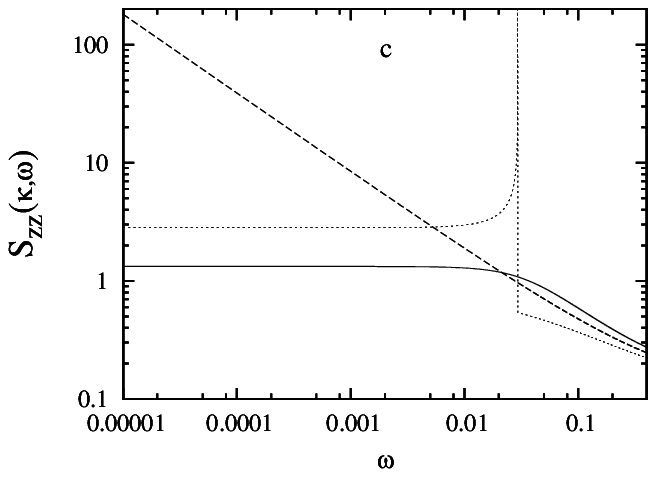, width = 0.32\linewidth}\\
\caption{Different kinds of potential van Hove singularities of the two-fermion dynamic quantities.
Frequency profiles of $S_{zz}(\kappa,\omega)$
(i.e. $S_{I}(\kappa,\omega)$ with $B_{I}(\kappa_1,\kappa_2)=1$)
at $T\to\infty$, $J=1$.
Panel a:
$\kappa=\pi/2$,
$K=0.25$ (dotted line),
$K=0.5$ (dashed line),
$K=2.5$ (solid line).
Panels b and c:
$K=2.5$,
$\kappa=2.70$ (dotted lines),
$\kappa=\kappa^\diamond\approx 2.7388 7681$ (dashed lines),
$\kappa=2.78$ (solid lines).
The initial slope of the dashed line in panel c corresponds to the dependence $\propto \omega^{-2/3}$.}
\label{fig05}
\end{figure}

Due to the presence of a two-particle density of states 
the two-fermion dynamic quantities (\ref{2.01}) and (\ref{2.04}) 
may exhibit a van Hove singularity
along the lines
\begin{eqnarray}
\omega_{s}(\kappa)
=
\left\{\Omega_{-}(\kappa), \Omega_{+}(\kappa)\right\},
\label{3.06}
\end{eqnarray}
or more precisely
along $\Omega_{-}(\kappa)$
[$\Omega_{+}(\kappa)$]
if $JK > 0$
[$JK < 0$]
for $\vert K/J\vert<1/2$
and
along both lines $\Omega_{-}(\kappa)$ and $\Omega_{+}(\kappa)$
for $\vert K/J\vert>1/2$.
Thus, a sufficiently strong three-site interaction $K$
increases the number of van Hove singularities.
This is nicely seen in Fig. \ref{fig04}
(and also in Fig. \ref{fig05})
where the two-fermion dynamic structure factors for $J=1$, $K=2.5$ at $T\to\infty$ are plotted.
[Recall that $B_{zz}(\kappa_1,\kappa_2)=1$
and therefore Fig. \ref{fig04}a with gray-scale plot for $S_{zz}(\kappa,\omega)$
most transparently demonstrates a new (low-frequency) line of van Hove singularities
emerging for $\vert K/J\vert>1/2$.]

Interestingly,
in addition to the conventional inverse square-root van Hove singularity
a singularity with exponent $-2/3$ may occur as $\vert K/J\vert>1/2$.
In fact,
plotting 
$\left.\partial^2 E(\kappa_1,\kappa_1+\kappa)/\partial \kappa_1^2\right\vert_{\kappa_1=\tilde{\kappa}_1}$ 
vs 
$\kappa$,
we note that 
$\left.\partial^2 E(\kappa_1,\kappa_1+\kappa)/\partial \kappa_1^2\right\vert_{\kappa_1=\tilde{\kappa}_1}=0$
for $\kappa=\kappa^\diamond$
which satisfies $x^+=1$, $JK>0$ (or $x^-=-1$, $JK<0$).
A similar analysis of 
$\left.\partial^3 E(\kappa_1,\kappa_1+\kappa)/\partial \kappa_1^3\right\vert_{\kappa_1=\tilde{\kappa}_1}$ 
vs 
$\kappa$
shows that
$\left.\partial^3 E(\kappa_1,\kappa_1+\kappa)/\partial \kappa_1^3\right\vert_{\kappa_1=\tilde{\kappa}_1}\ne 0$
for $\kappa=\kappa^\diamond$.
Moreover,
we find that $E(\kappa_1,\kappa_1+\kappa)=0$
for these values of $\kappa$ and $\kappa_1$.
This immediately implies that
$S_{I}(\kappa^\diamond,\epsilon)\propto\epsilon^{-2/3}$,
$\epsilon\to +0$.
In Fig. \ref{fig05}
we demonstrate potential van Hove singularities of the two-fermion dynamic structure factors.
In particular,
we illustrate the van Hove singularity with exponent $-2/3$ for $J=1$, $K=2.5$.
Solving equation $x^+=1$ (\ref{3.03}) with respect to $\kappa$
we find $\kappa^\diamond\approx 2.7388 7681$.
The frequency profiles around this value of wavevector
clearly show two types of van Hove singularity,
i.e. with exponent $-1/2$ 
(in most cases when $\omega\to\omega_s(\kappa)-\epsilon$;
e.g., the dotted and solid lines in Figs. \ref{fig05}b and \ref{fig05}c)
and with exponent $-2/3$ 
(only when $\kappa=\kappa^\diamond$ and $\omega\to \epsilon$;
the dashed lines in Figs. \ref{fig05}b and \ref{fig05}c).

We consider now the case of zero temperature $T=0$
when the Fermi functions 
entering the function $C(\kappa_1,\kappa_2)$ in Eq. (\ref{2.04})
become extremely important.
They imply that
in the ground state we have to require in addition to Eq. (\ref{3.01})
$\Lambda_{\kappa_1}\le 0$
and
$\Lambda_{\kappa_2}\ge 0$.
We first consider the case $\vert K/J\vert <1/2$.
Plotting the dependence of $E(\kappa_1,\kappa_1+\kappa)C(\kappa_1,\kappa_1+\kappa)$ on $\kappa_1$
we find
that the two characteristic curves,
$\Lambda_{\kappa_1+\kappa}$ with $\kappa_1$ satisfying $\Lambda_{\kappa_1}=0$
(or $E(\kappa_1,\kappa_1+\kappa)$ with $\kappa_1$ satisfying $\Lambda_{\kappa_1}=0$)
and
$-\Lambda_{\kappa_1}$ with $\kappa_1$ satisfying $\Lambda_{\kappa_1+\kappa}=0$
(or $E(\kappa_1,\kappa_1+\kappa)$ with $\kappa_1$ satisfying $\Lambda_{\kappa_1+\kappa}=0$),
play a special role.
Solving the equation $\Lambda_{k}=0$ for $k$, or more precisely for $y=\cos k$,
we find
\begin{eqnarray}
y^{\pm}
=
\frac{J}{2K}\pm\sqrt{\frac{J^2}{4K^2}+\frac{\Omega}{K}+\frac{1}{2}}.
\label{3.07}
\end{eqnarray}
Taking into account that $\vert y\vert\le 1$,
we see that for $\vert K/J\vert <1/2$,
Eq. (\ref{3.07}) may yield two $k$ values,
$\check{k}=\arccos y^{-}$, $\check{k}=-\arccos y^{-}$ (for $JK>0$),
or
$\check{k}=\arccos y^{+}$, $\check{k}=-\arccos y^{+}$ (for $JK<0$).
For $\vert K/J\vert >1/2$ there may be two such pairs of $k$, 
$\check{k}=\arccos y^{-}$, $\check{k}=-\arccos y^{-}$
and
$\check{k}=\arccos y^{+}$, $\check{k}=-\arccos y^{+}$;
we will discuss that case later.

\begin{figure}
\epsfig{file = 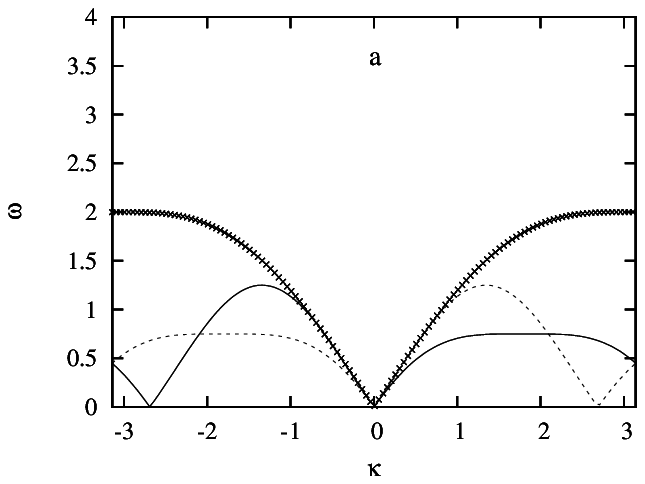, width = 0.32\linewidth}
\epsfig{file = 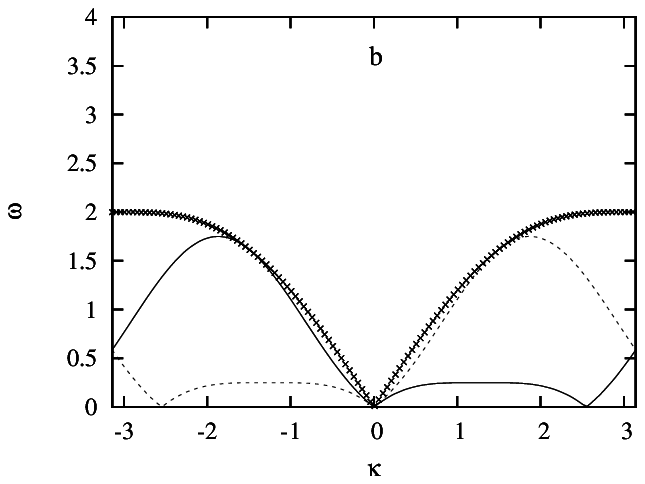, width = 0.32\linewidth}
\epsfig{file = 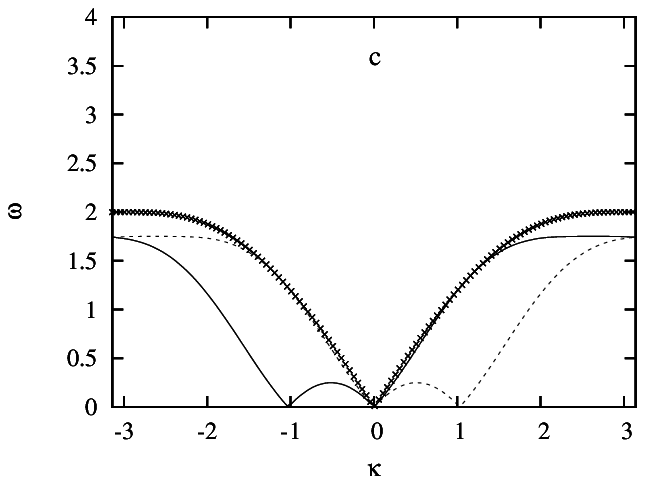, width = 0.32\linewidth}\\
\epsfig{file = 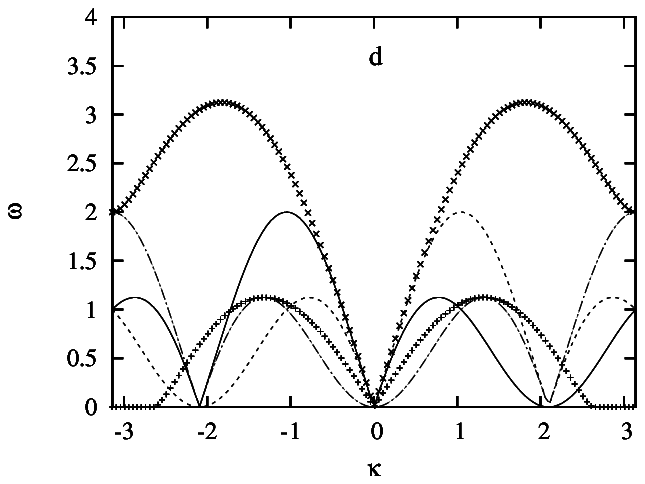, width = 0.32\linewidth}
\epsfig{file = 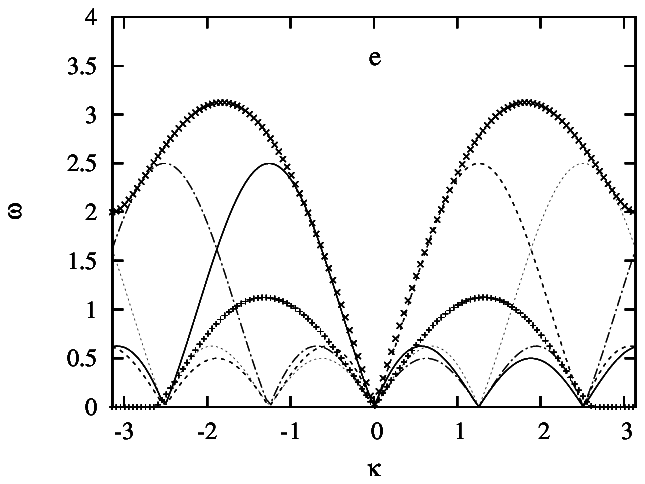, width = 0.32\linewidth}
\epsfig{file = 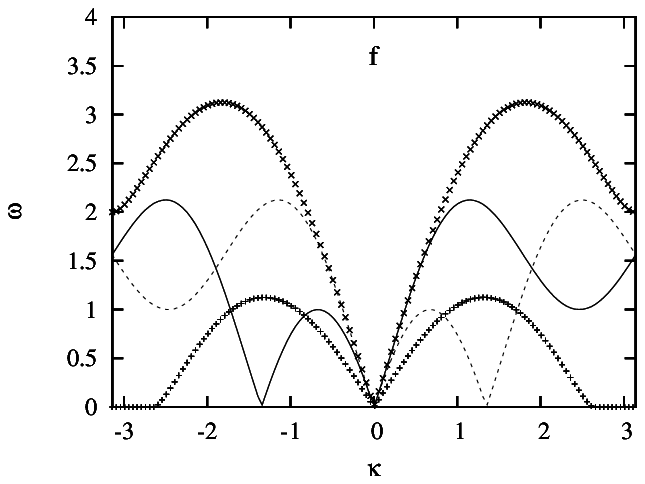, width = 0.32\linewidth}\\
\epsfig{file = 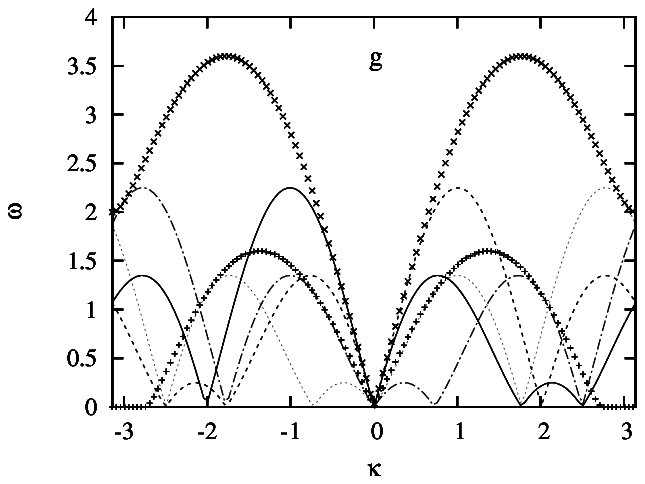, width = 0.32\linewidth}
\epsfig{file = 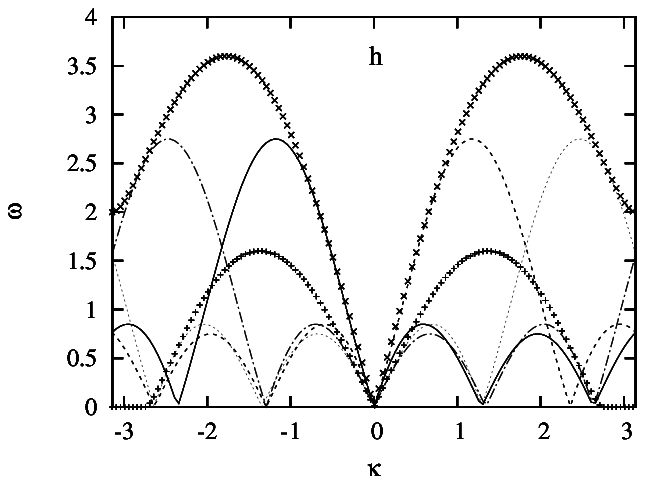, width = 0.32\linewidth}
\epsfig{file = 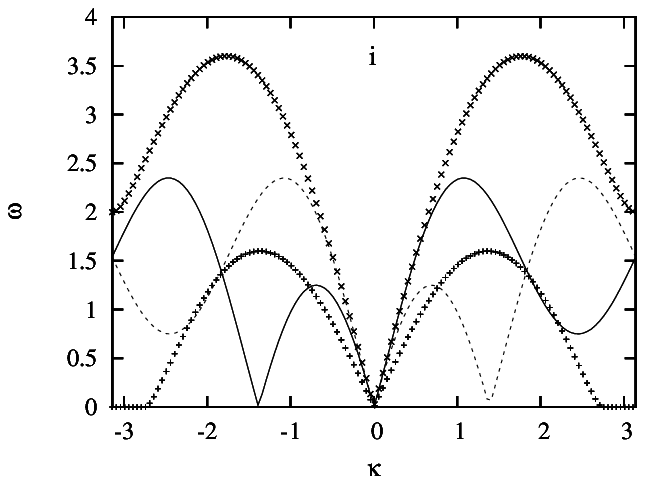, width = 0.32\linewidth}
\caption{Characteristic lines of the two-fermion excitation continuum.
We assume $J=1$
and
$K=0.5$, $\Omega=0$, $\Omega=-0.5$, $\Omega=1$ (panels a, b, c),
$K=2$,   $\Omega=0$, $\Omega=-0.5$, $\Omega=1$ (panels d, e, f),
$K=2.5$, $\Omega=0$, $\Omega=-0.5$, $\Omega=1$ (panels g, h, i).
{\sf{x}} and {\sf{+}} symbols correspond to $\Omega_{-}(\kappa)$ and $\Omega_{+}(\kappa)$, respectively.
Solid, dashed, dash-dotted and dotted lines correspond to 
$\omega_{-}^{+}(\kappa)$,
$\omega_{-}^{-}(\kappa)$,
$\omega_{+}^{+}(\kappa)$
and
$\omega_{+}^{-}(\kappa)$,
respectively.
(The panels a, d, g 
can be compared to the panels a, b, c in Fig. \ref{fig02}.)}
\label{fig06}
\end{figure}

For $\vert K/J\vert <1/2$ we consider two characteristic lines 
\begin{eqnarray}
\omega_{-}^{+}(\kappa)=\vert \Lambda_{\arccos y^{-}+\kappa}\vert,
\;\;\;
\omega_{-}^{-}(\kappa)=\vert \Lambda_{-\arccos y^{-}+\kappa}\vert,
\;\;\;
JK>0
\label{3.08}
\end{eqnarray}
or
\begin{eqnarray}
\omega_{+}^{+}(\kappa)=\vert \Lambda_{\arccos y^{+}+\kappa}\vert,
\;\;\;
\omega_{+}^{-}(\kappa)=\vert \Lambda_{-\arccos y^{+}+\kappa}\vert,
\;\;\;
JK<0.
\label{3.09}
\end{eqnarray}
The smaller one of the two values 
$\omega_{-}^{i}(\kappa)$, $i=-,+$ for $JK>0$  
[$\omega_{+}^{i}(\kappa)$, $i=-,+$ for $JK<0$]
gives the lower boundary of the ground-state two-fermion excitation continuum $\omega_l(\kappa)$,
whereas the other (larger) one gives 
either 
the upper boundary of the ground-state two-fermion excitation continuum $\omega_u(\kappa)$
or
the middle boundary of the ground-state two-fermion excitation continuum $\omega_m(\kappa)$.
The former case occurs if $\tilde{\kappa}_1$ 
yielding $\left.\partial E(\kappa_1,\kappa_1+\kappa)/\partial\kappa_1\right\vert_{\kappa_1=\tilde{\kappa}_1}=0$ 
(see above)
belongs to the region of $\kappa_1$ 
where $E(\kappa_1,\kappa_1+\kappa)C(\kappa_1,\kappa_1+\kappa)=0$
(as, e.g., is seen in Fig. \ref{fig06}b for small $\vert\kappa\vert$).
In the latter case, 
when $\tilde{\kappa}_1$ 
yielding $\left.\partial E(\kappa_1,\kappa_1+\kappa)/\partial\kappa_1\right\vert_{\kappa_1=\tilde{\kappa}_1}=0$ 
does belong to the region of $\kappa_1$ 
where $E(\kappa_1,\kappa_1+\kappa)C(\kappa_1,\kappa_1+\kappa)\ne 0$,
the larger value of
$\omega_{-}^{i}(\kappa)$, $JK>0$  (\ref{3.08})
[$\omega_{+}^{i}(\kappa)$, $JK<0$ (\ref{3.09})] 
gives the middle boundary of the ground-state two-fermion excitation continuum $\omega_m(\kappa)$ 
whereas the upper boundary $\omega_u(\kappa)$ is given by Eq. (\ref{3.04}) [Eq. (\ref{3.05})].
For the frequencies $\omega$ between $\omega_l(\kappa)$ and $\omega_m(\kappa)$
[$\omega_m(\kappa)$ and $\omega_u(\kappa)$] 
equation (\ref{2.02}) has 
one solution 
[two solutions] 
$\kappa^{\star}_1$ .
Thus,
the ground-state $S_{I}(\kappa,\omega)$ changes by a factor 2 
at the middle boundary $\omega_m(\kappa)$.

The soft modes $\kappa_0$ can be determined from the equation 
$\omega_{-}^{\pm}(\kappa_0)=0$, $JK>0$
[$\omega_{+}^{\pm}(\kappa_0)=0$, $JK<0$].
Therefore,
if $JK>0$,
\begin{eqnarray}
\kappa_0=\left\{0,\pm 2\arccos y^-\right\},
\;\;\;
y^->0,
\nonumber\\
\kappa_0=\left\{0,\mp 2\arccos y^-\pm 2\pi\right\},
\;\;\;
y^-<0
 \label{3.10}
\end{eqnarray}
or,
if $JK<0$,
\begin{eqnarray}
\kappa_0=\left\{0,\pm 2\arccos y^+\right\},
\;\;\;
y^+>0,
\nonumber\\
\kappa_0=\left\{0,\mp 2\arccos y^+\pm 2\pi\right\},
\;\;\;
y^+<0.
\label{3.11}
\end{eqnarray}

We now turn to the case $\vert K/J\vert >1/2$. 
As already mentioned,
equation (\ref{3.07}) may yield two pairs of $k$,
$\check{k}=\arccos y^{-}$, $\check{k}=-\arccos y^{-}$
and
$\check{k}=\arccos y^{+}$, $\check{k}=-\arccos y^{+}$,
and therefore all four characteristic lines 
in the $\kappa$--$\omega$ plane,
$\omega_{-}^{+}(\kappa)$,
$\omega_{-}^{-}(\kappa)$,
$\omega_{+}^{+}(\kappa)$,
$\omega_{+}^{-}(\kappa)$
given by Eqs. (\ref{3.08}), (\ref{3.09})
come into play simultaneously.
Thus, 
as $\vert K\vert$ exceeds $\vert J\vert/2$ an ``extra'' ground-state two-fermion excitation continuum emerges.
Its lower boundary and upper/middle boundary are given by formulas (\ref{3.08}), (\ref{3.09})
[in the case when Eqs. (\ref{3.08}), (\ref{3.09}) give the middle boundary, 
the upper boundary is given by one of the formulas in Eqs. (\ref{3.04}), (\ref{3.05})].
The number of soft modes increases but cannot exceed 9.

In the vicinity of a soft mode $\kappa_0$ 
the lower boundary of the two-fermion continuum in most cases displays a ``{\sf{V}}'' shape,
i.e. it is proportional to $\vert\kappa-\kappa_0\vert$.
However,
it is worth noting 
that a parabolic shape $\propto(\kappa-\kappa_0)^2$ is also possible 
for suitable parameter combinations. 
To see this we recall that the lower boundary is basically determined by the equation
$\Lambda_k=0$ with $\Lambda_k=J\cos k-(K/2)\cos(2k)+\Omega$.
For the conventional $XX$ chain ($K=0$) in a transverse field $\Omega$ 
we see that at the critical field values $\Omega=\pm\vert J\vert$ 
the dispersion $\Lambda_k\propto (k-k_0)^2$, 
with a corresponding parabolic shape of the lower continuum boundary 
near a soft mode at $\kappa_0$.
[This can be seen, e.g., in Fig. 11f of Ref. \cite{dks}.]
At the critical field the nature of the ground state of the $XX$ chain changes 
from partially filled to completely filled or completely empty 
in terms of Jordan-Wigner fermions.
For nonzero $K$ 
the $\cos(2k)$-term may generate additional maxima or minima in $\Lambda_{k}$.
That implies the emergence of additional critical field values,
at which $k$ regions near the additional minima or maxima of $\Lambda_{k}$ 
open or close for occupation by Jordan-Wigner fermions.
These critical values correspond to the lines 
separating different ground-state phases in Fig. \ref{fig01}. 
Along these lines we expect parabolic behavior 
of the lower two-fermion continuum boundary 
(see, e.g., Figs. \ref{fig02}b, \ref{fig02}d, \ref{fig02}e, \ref{fig02}f).

Summarizing this part,
we report in Fig. \ref{fig06} all the characteristic lines 
of the two-fermion excitation continuum discussed above.
In these plots
symbols correspond to $\Omega_{-}(\kappa)$ ({\sf{x}} symbols) and $\Omega_{+}(\kappa)$ ({\sf{+}} symbols),
whereas lines correspond to 
$\omega_{-}^{+}(\kappa)$ (solid),
$\omega_{-}^{-}(\kappa)$ (dashed),
$\omega_{+}^{+}(\kappa)$ (dash-dotted),
$\omega_{+}^{-}(\kappa)$ (dotted).
For $\vert K/J\vert <1/2$ only three characteristic lines are relevant,
but
if $\vert K/J\vert >1/2$ all six lines are relevant.
These lines are important not only for understanding the distribution of the two-fermion dynamic structure factors
$S_{I}(\kappa,\omega)$ (\ref{2.01}), (\ref{2.04}) over the $\kappa$ -- $\omega$ plane
(see gray-scale plots in Figs. \ref{fig02}, \ref{fig03}, \ref{fig04})
but also for many-fermion dynamic structure factors
(like $S_{xx}(\kappa,\omega)$)
at low temperatures
as will be discussed in the next section
(see gray-scale plots in Fig. \ref{fig10}).

Finally,
we discuss some specific properties of the two-fermion dynamic structure factors (\ref{2.04})
controlled by different $B$-functions.
Comparing different panels in Figs. \ref{fig03} and \ref{fig04} 
we observe a number of small but definite differences 
for the detailed distributions of $S_{I}(\kappa,\omega)$ 
over the $\kappa$ -- $\omega$ plane.

\begin{figure}
\epsfig{file = 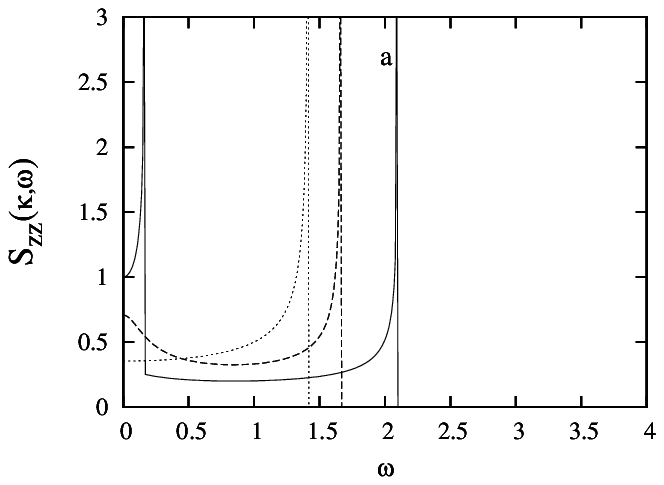, width = 0.32\linewidth}
\epsfig{file = 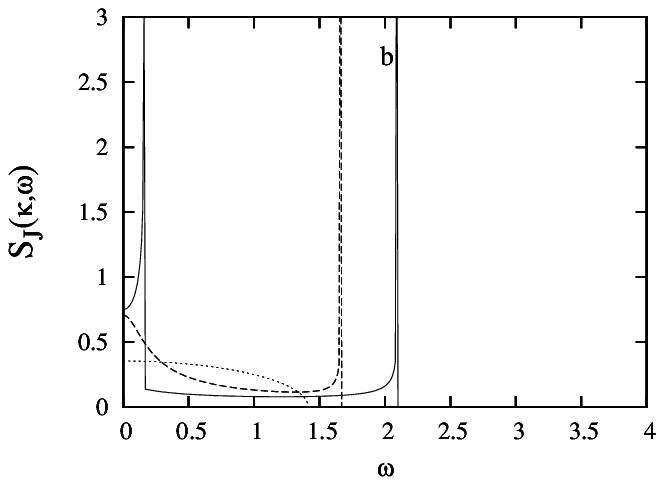, width = 0.32\linewidth}
\epsfig{file = 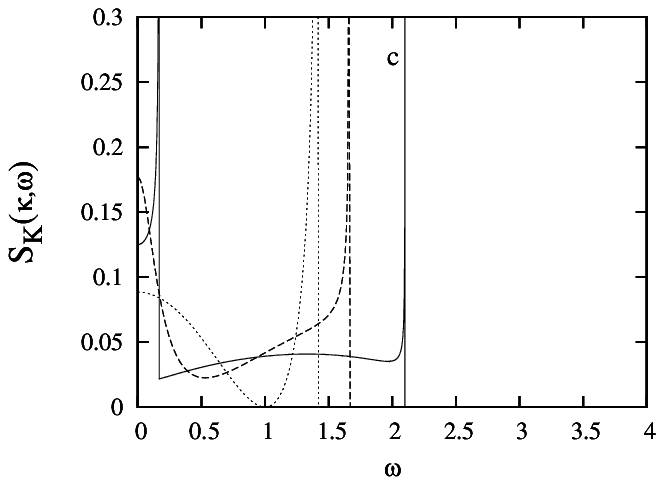, width = 0.32\linewidth}
\caption{Illustrating the role of $B$-functions.
$S_{zz}(\pi/2,\omega)$ (panel a),
$S_{J}(\pi/2,\omega)$ (panel b),
$S_{K}(\pi/2,\omega)$ (panel c)
at $T\to\infty$.
$J=1$,
$K=0$ (dotted lines),
$K=0.5$ (dashed lines),
$K=1$ (solid lines).}
\label{fig07}
\end{figure}

To be specific,
we may focus on the dynamic dimer structure factor $S_{J}(\kappa,\omega)$. 
It is known
that $S_{J}(\kappa,\omega)$ does not diverge along the upper boundary
owing to $B_{J}(\kappa_1,\kappa_2)$ for the conventional $XX$ chain,
i.e. when $K=0$
(see, e.g., Ref. \onlinecite{dksm} and references therein).
This can be also seen in Fig. \ref{fig07}b
where the dotted line corresponds to $K=0$.
This changes, however, if $K\ne 0$:
the dynamic dimer structure factor exhibits a van Hove singularity along the upper boundary
(the dashed and solid lines in  Fig. \ref{fig07}b)
indicating the presence of nonzero three-site interactions.

Next we may consider $S_{K}(\kappa,\omega)$.
As can be seen in Figs. \ref{fig04}c and \ref{fig07}c, 
the van Hove singularity at the upper boundary becomes less distinctive as $K$ increases.
To explain this, 
we introduce the notation 
$x=\cos(\kappa/2+\kappa_1)$
and rewrite $\partial E(\kappa_1,\kappa_1+\kappa)/\partial\kappa_1$ (\ref{3.02}) 
as
$2\sin(\kappa/2)(2K\cos(\kappa/2)(2x^2-1)-Jx)$,
whereas $B_{K}(\kappa_1,\kappa_1+\kappa)$ (\ref{2.04})
as
$(1/4)(2x^2-1)^{2}$. 
In the vicinity of the upper boundary 
the denominator in Eq. (\ref{2.04}) for $S_{K}(\kappa,\omega)$ tends to zero,
however,
in the limit $K\to\infty$ 
the numerator in Eq. (\ref{2.04}) for $S_{K}(\kappa,\omega)$ becomes proportional to the denominator squared 
which makes the fraction equal to zero.
Thus, for any finite large $K$ the van Hove singularity at the upper boundary does exist 
(although with increasing $K$ it is harder to find it numerically)
and it disappears only in the limit $K\to\infty$.

\section{Many-fermion dynamic quantities}
\label{sec4}
\setcounter{equation}{0}

In this section we discuss many-fermion dynamic quantities
fixing for concreteness our attention 
to the $xx$ dynamic structure factor $S_{xx}(\kappa,\omega)$. 
First we report two analytical results 
referring to the high-temperature limit and to the zero-temperature strong-field regime,
respectively, 
and then we turn to high precision numerical data 
for arbitrary values of temperature and the Hamiltonian parameters.

We first consider 
the $xx$ two-spin time-dependent correlation function $\langle s_j^x(t)s_{j+n}^x\rangle$ 
at  $T\to\infty$.
Since the Zeeman term commutes with the Hamiltonian of the considered model (\ref{1.01})
and in the high-temperature limit $\exp(-\beta H)\to 1$ 
and consequently the averages of spin operators are zero 
we can easily extract the dependence on the transverse field $\Omega$
\begin{eqnarray}
\label{4.01}
\langle s_j^x(t)s_{j+n}^x\rangle
=
\cos\left(\Omega t\right)
\left.\langle s_j^x(t)s_{j+n}^x\rangle\right\vert_{\Omega=0}.
\end{eqnarray}
Thus we can proceed assuming $\Omega=0$.
Next we substitute into the spin correlation function on the r.h.s. in  Eq. (\ref{4.01}) 
a short-time expansion 
$s^x_j(t)=s^x_j+{\rm{i}}\left[H,s_j^x\right]t
-(1/2)\left[H,\left[H,s_j^x\right]\right]t^2+\ldots$ \cite{dls}
and after simple but tedious calculations find
\begin{eqnarray}
\label{4.02}
4\left.\langle s_j^x(t)s_{j+n}^x\rangle\right\vert_{\Omega=0}
=
\delta_{n,0}
\left(
1-\left(\frac{J^2_{j-1}+J^2_{j}}{8}+\frac{K^2_{j-2}+2K^2_{j-1}+K^2_{j}}{32}\right)t^2+\ldots
\right),
\end{eqnarray}
where for generality we have considered a model with position-dependent couplings.
Eq. (\ref{4.02}) is consistent with the Gaussian decay
\begin{eqnarray}
\label{4.03}
4\langle s_j^x(t)s_{j+n}^x\rangle
=
\delta_{n,0}
\cos\left(\Omega t\right)
\exp\left(-\left(\frac{J^2}{4}+\frac{K^2}{8}\right)t^2\right)
\end{eqnarray}
for the model with position-independent couplings.
Using MAPLE codes 
we checked that the terms up to $t^4$ in the short-time expansion for $\langle s_j^x(t)s_{j+n}^x\rangle$
indeed agree with Eq. (\ref{4.03}).

Alternative (although not independent) arguments supporting Eq. (\ref{4.03}) 
follow Refs. \onlinecite{florencio,stolze}.
We examine the continued-fraction coefficients $\Delta_k$ of the relaxation function
$c^{xx}(z)=4\int_0^{\infty}{\rm{d}}t\exp(-zt)\langle s_j^x(t)s_j^x\rangle
=1/(z+\Delta_1/(z+\Delta_2/(z+\ldots)))$ 
at $T\to\infty$.
The sequence of the continued-fraction coefficients $\Delta_k$ 
reflects the time dependence of the associated autocorrelation function,
and, 
in particular, 
when $\Delta_k= k\Delta$ 
then $4\langle s_j^x(t)s_j^x\rangle=\exp(-\Delta t^2/2)$.
The sequence $\Delta_k$ can be determined by the methods elaborated in Ref. \onlinecite{stolze}.
We designed a MAPLE program which in a reasonable amount of time calculated $\Delta_k$ for $k=1,2,3,4$
and confirmed the Gaussian decay (\ref{4.03}).

Finally,
from our calculations we also find a more general result for the homogeneous model (\ref{1.01}) 
which is given by Eq. (\ref{4.03}) after the substitution
$J^2\to J^2+D^2$, 
$K^2\to K^2+E^2$.

\begin{figure}
\epsfig{file = 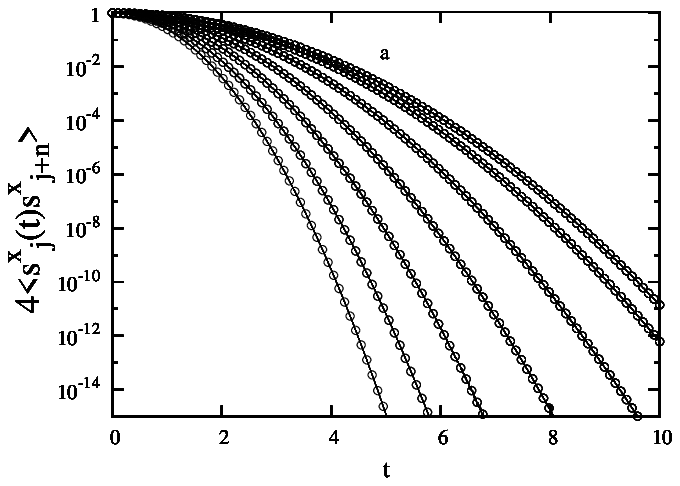, width = 0.32\linewidth}\\
\epsfig{file = 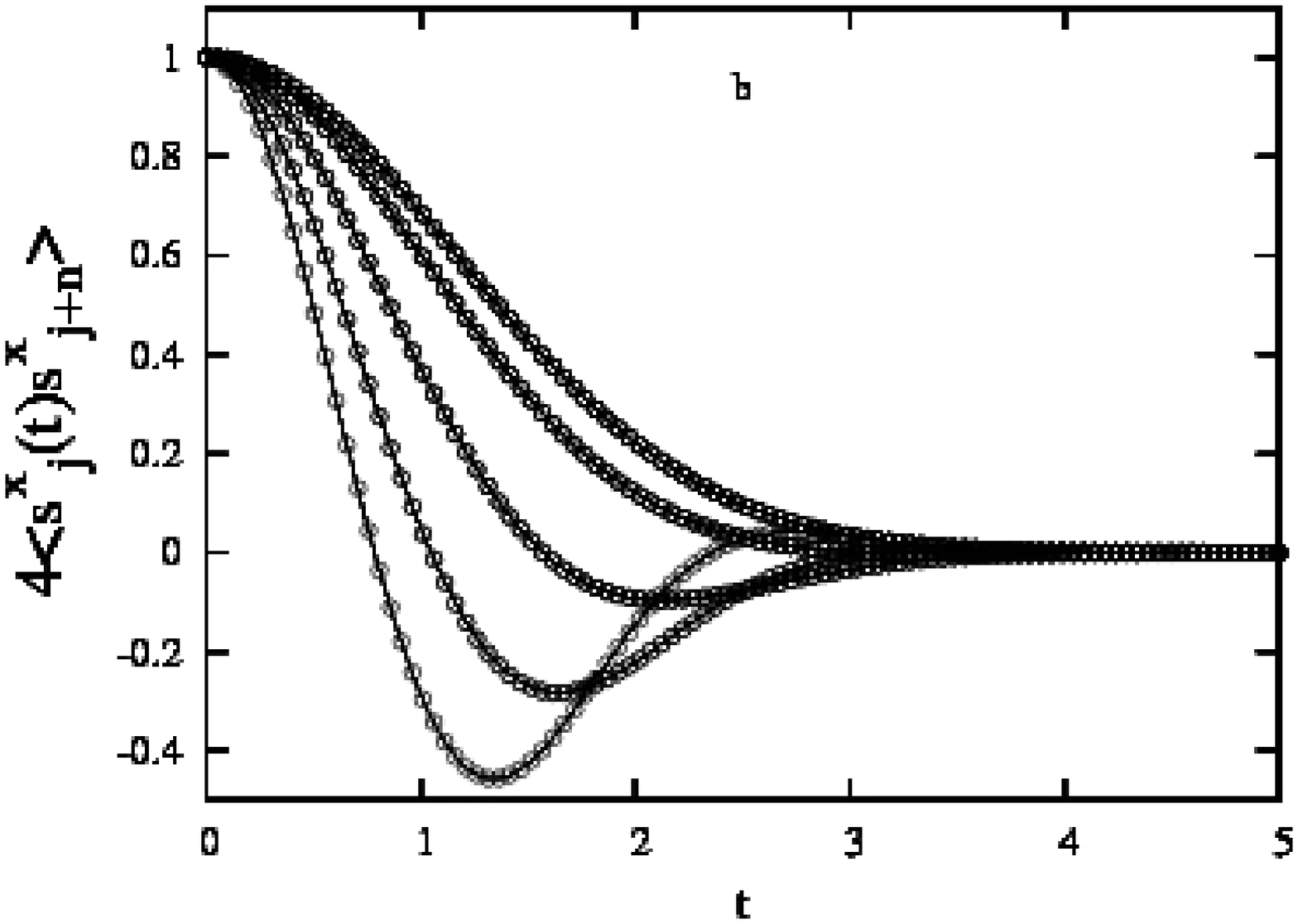, width = 0.32\linewidth}
\caption{Time dependence of the $xx$ autocorrelation function
in the high-temperature limit ($\beta=0.0001$); $J=1$.
Panel a:
$\langle s_j^x(t)s_j^x\rangle$ at $\Omega=0$, $K=0,\;0.5,\;1,\;1.5,\;2,\;2.5,\;3$ (from top to bottom).
Panel b:
$\langle s_j^x(t)s_j^x\rangle$ at $K=1$, $\Omega=0,\;0.5,\;1,\;1.5,\;2$ (from top to bottom).
Symbols correspond to Eq. (\ref{4.03});
lines correspond to numerical data obtained for $N=400$, $j=41$.}
\label{fig08}
\end{figure}

To summarize,
in Fig. \ref{fig08} 
we compare analytical predictions according to Eq. (\ref{4.03}) (symbols)
with numerical calculations (lines) (see below)
and observe an excellent agreement between both sets of data. 
Eq. (\ref{4.03}) provides an extension of the well-known result 
for the conventional transverse $XX$ chain \cite{inf_temp} 
for the kind of three-site interactions considered here.
The presented arguments in favor of the Gaussian decay (\ref{4.03})
may be put even on a more rigorous foundation 
using the approach elaborated in Ref.~\onlinecite{perk_capel_1978}.

\begin{figure}
\epsfig{file = 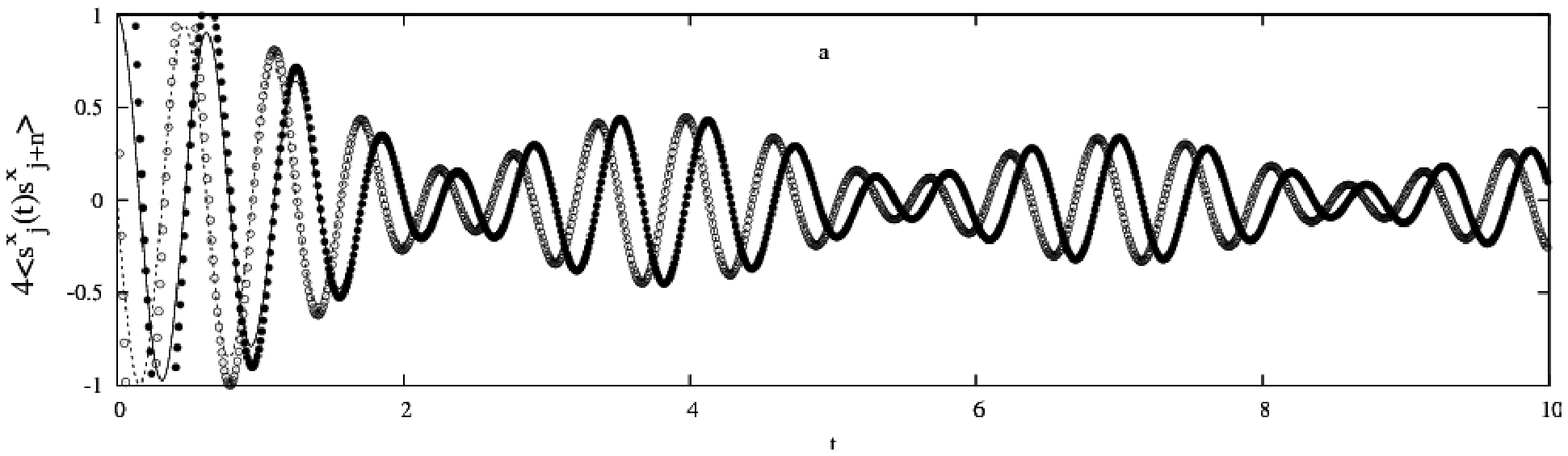, width = 0.64\linewidth}\\
\epsfig{file = 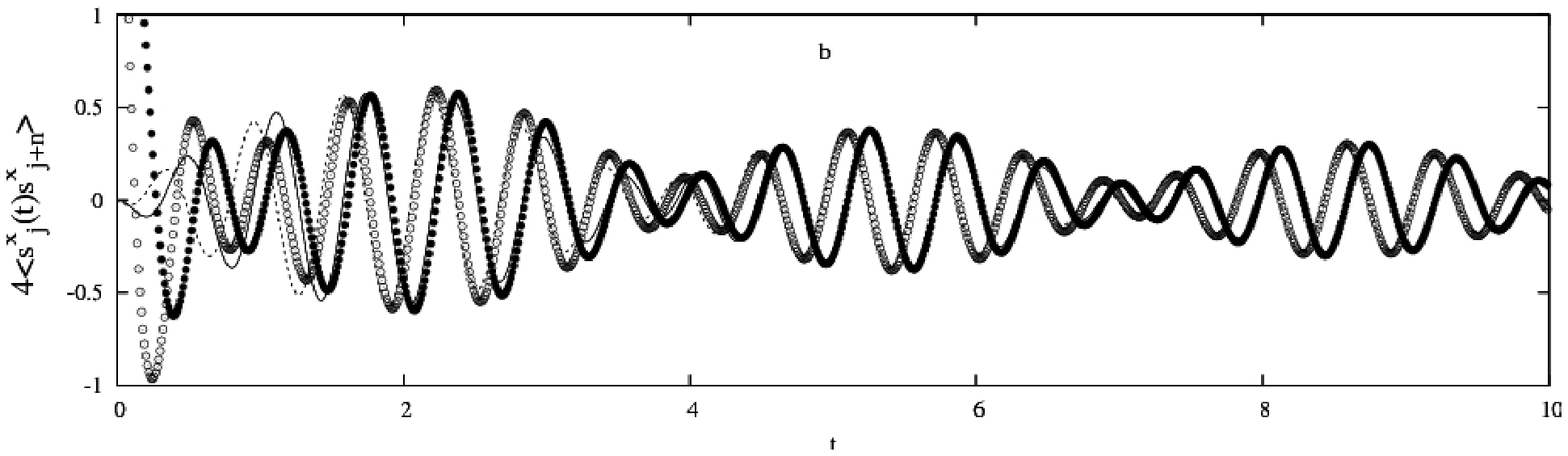, width = 0.64\linewidth}
\caption{Time dependence of
$\langle s_j^x(t)s_{j+n}^x\rangle$
(panel a: $n=0$;
panel b: $n=1$)
in the low-temperature limit ($\beta=100$); $J=1$, $K=0.25$, $\Omega=10$.
Lines correspond to numerical data for $N=400$, $j=101$.
Symbols correspond to the long-time asymptotics given by Eq. (\ref{a.06}).
(We notice that the asymptotic becomes accurate already for short times.)
Solid (dashed) lines or filled (empty) symbols refer to real (imaginary) part of $\langle s_j^x(t)s_{j+n}^x\rangle$.}
\label{fig09}
\end{figure}

Next we turn to the zero-temperature strong-field regime.
More precisely, we consider the ferromagnetic phase (light) in Fig. \ref{fig01}. 
In the ferromagnetic phase the ground state of the spin model is completely polarized,
i.e.
$\vert{\rm{GS}}\rangle=\prod_{n}\vert\downarrow\rangle_n$
($\vert{\rm{GS}}\rangle=\prod_{n}\vert\uparrow\rangle_n$)
for positive (negative) $\Omega$,
that permits easily to take into account the Jordan-Wigner sign factors 
entering the formula for $\langle s_j^x(t)s_{j+n}^x\rangle$ \cite{zero_temp}.
Assuming, for example, 
$\vert{\rm{GS}}\rangle=\prod_{n}\vert\downarrow\rangle_n$
(light region in the upper half-plane in Fig. \ref{fig01})
we immediately get
\begin{eqnarray}
\label{4.04}
4\langle s_j^x(t)s_{j+n}^x\rangle
=
\frac{1}{N}
\sum_{\kappa}\exp\left({\rm{i}}\left(\kappa n-\Lambda_{\kappa}t\right)\right)
\nonumber\\
\stackrel{N \rightarrow \infty}{\to}
\frac{1}{2\pi}\int_{-\pi}^{\pi}{\rm{d}}\kappa
\exp\left({\rm{i}}\left(\kappa n-\Lambda_{\kappa}t\right)\right)
\end{eqnarray}
with $\Lambda_{\kappa}=J\cos\kappa-(K/2)\cos\left(2\kappa\right)+\Omega>0$.
For 
$\vert{\rm{GS}}\rangle=\prod_{n}\vert\uparrow\rangle_n$
(light region in the lower half-plane in Fig. \ref{fig01})
we have
\begin{eqnarray}
\label{4.05}
4\langle s_j^x(t)s_{j+n}^x\rangle
=
\frac{1}{N}
\sum_{\kappa}\exp\left(-{\rm{i}}\left(\left(\kappa\mp\pi\right)n-\Lambda_{\kappa}t\right)\right)
\nonumber\\
\stackrel{N \rightarrow \infty}{\to}
\frac{1}{2\pi}\int_{-\pi}^{\pi}{\rm{d}}\kappa
\exp\left(-{\rm{i}}\left(\left(\kappa\mp\pi\right)n-\Lambda_{\kappa}t\right)\right)
\end{eqnarray}
with $\Lambda_{\kappa}=J\cos\kappa-(K/2)\cos\left(2\kappa\right)+\Omega<0$.
Eqs. (\ref{4.04}), (\ref{4.05}) contain the result 
for the conventional transverse $XX$ chain \cite{zero_temp},
$4\langle s_j^x(t)s_{j+n}^x\rangle
=\exp(-{\rm{i}}\vert\Omega\vert t)(-{\rm{i}})^nJ_{n}(Jt)$,
where $J_n(z)$ is the Bessel function of the first kind \cite{be2}.
Some further properties of $\langle s_j^x(t)s_{j+n}^x\rangle$ are collected in the Appendix.
We only notice here
that in the regime considered the $xx$ time-dependent correlation function oscillates,
with the envelope decaying proportional to $t^{-1/2}$ as $t\to\infty$
for $\vert K/J\vert\ne 1/2$ 
or 
proportional to $t^{-1/4}$ for $\vert K/J\vert = 1/2$
(see Eqs. (\ref{a.06}), (\ref{a.07}), (\ref{a.08})
and Fig. \ref{fig09}).

Eqs. (\ref{4.03}) and (\ref{4.04}), (\ref{4.05}) immediately yield the $xx$ dynamic structure factor (\ref{1.08}).
In the high-temperature limit we have
\begin{eqnarray}
\label{4.06}
S_{xx}(\kappa,\omega)
=
\frac{\sqrt{\pi}}{4\sqrt{J^2+\frac{K^2}{2}}}
\left(
\exp\left(-\frac{\left(\omega-\Omega\right)^2}{J^2+\frac{K^2}{2}}\right)
+
\exp\left(-\frac{\left(\omega+\Omega\right)^2}{J^2+\frac{K^2}{2}}\right)
\right),
\end{eqnarray}
i.e. the $xx$ dynamic structure factor displays 
a $\kappa$-independent Gaussian ridge 
centered at frequency $\vert\Omega\vert$ with the width controlled by interspin interactions.
In the zero-temperature strong-field regime we have
\begin{eqnarray}
\label{4.07}
S_{xx}(\kappa,\omega)
=
\frac{\pi}{2}\delta\left(\omega-\omega^{\star}(\kappa)\right),
\nonumber\\
\omega^{\star}(\kappa)
=
\left\{
\begin{array}{ll}
\vert\Omega\vert+J\cos\kappa-\frac{K}{2}\cos(2\kappa), & \Omega>0, \\
\vert\Omega\vert+J\cos\kappa+\frac{K}{2}\cos(2\kappa), & \Omega<0,
\end{array}
\right.
\end{eqnarray}
i.e. the $xx$ dynamic structure factor displays 
a $\delta$-peak 
along the line $\omega^{\star}(\kappa)$ (\ref{4.07}) in the $\kappa$ -- $\omega$ plane.
Interestingly,
if $K\ne 0$ the symmetry of Eq. (\ref{4.07}) with respect to the change $\Omega\to -\Omega$ is broken, 
in agreement with the ground-state phase diagram shown in Fig. \ref{fig01}.

\begin{figure}
\epsfig{file = 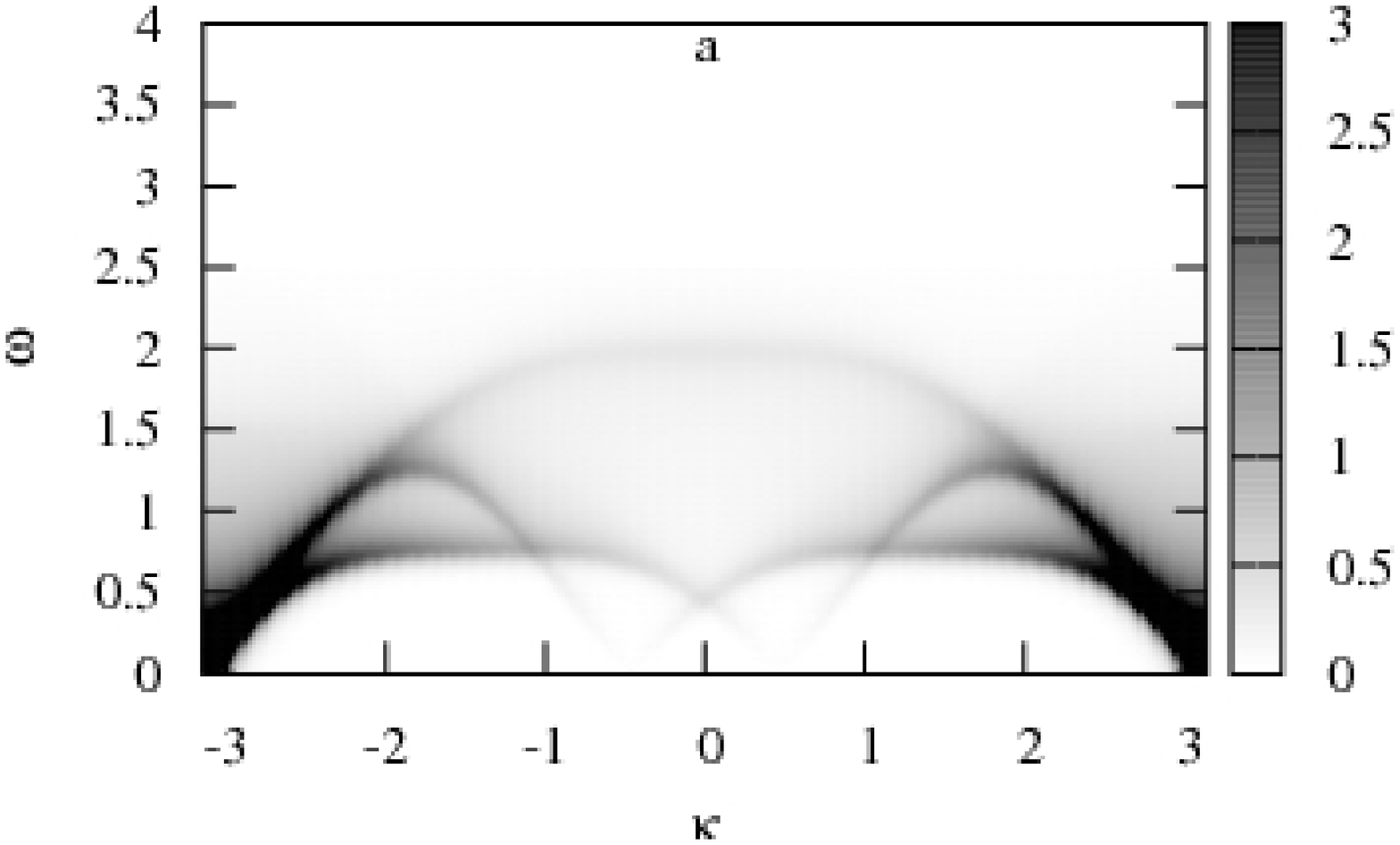, width = 0.32\linewidth}
\epsfig{file = 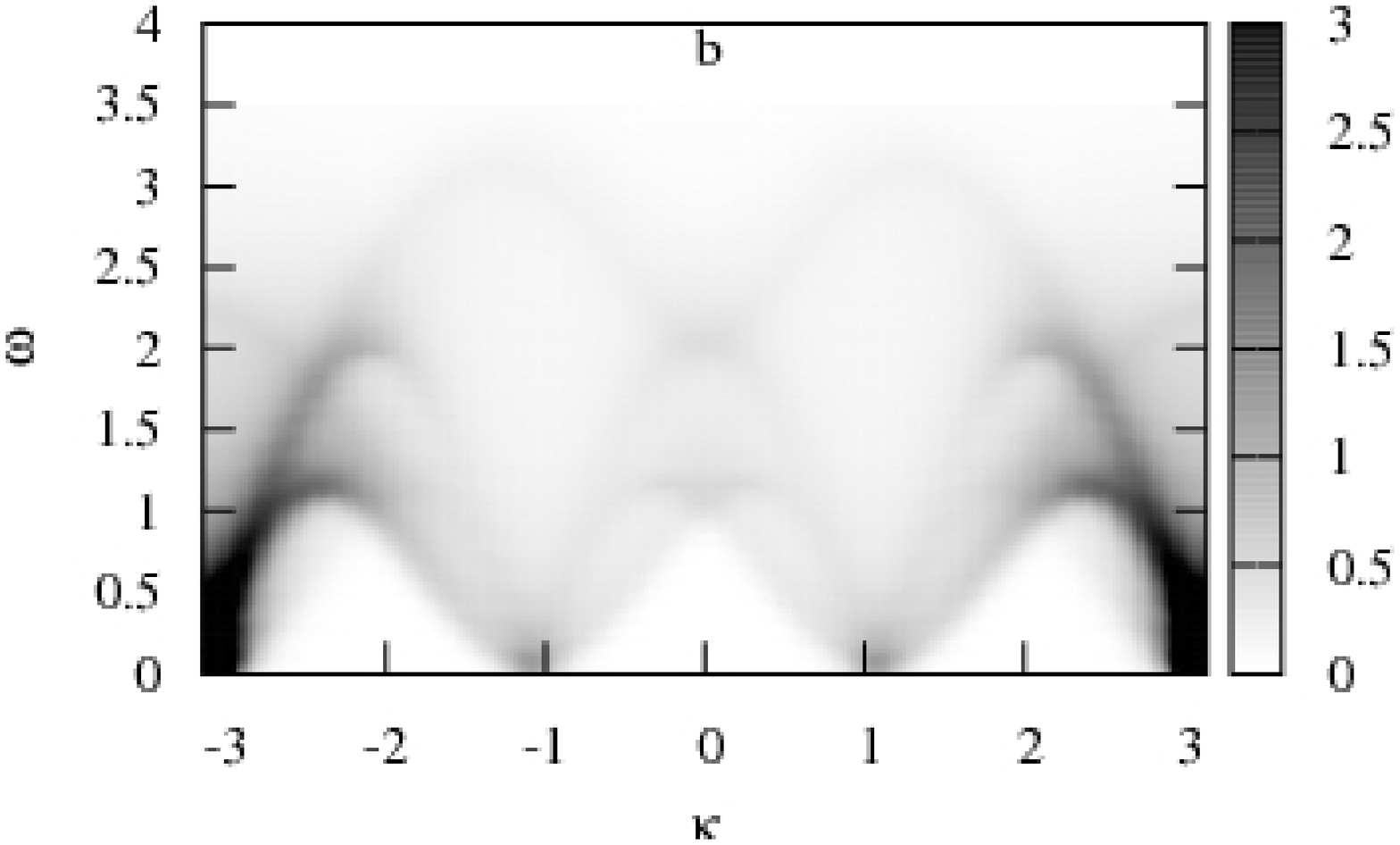, width = 0.32\linewidth}
\epsfig{file = 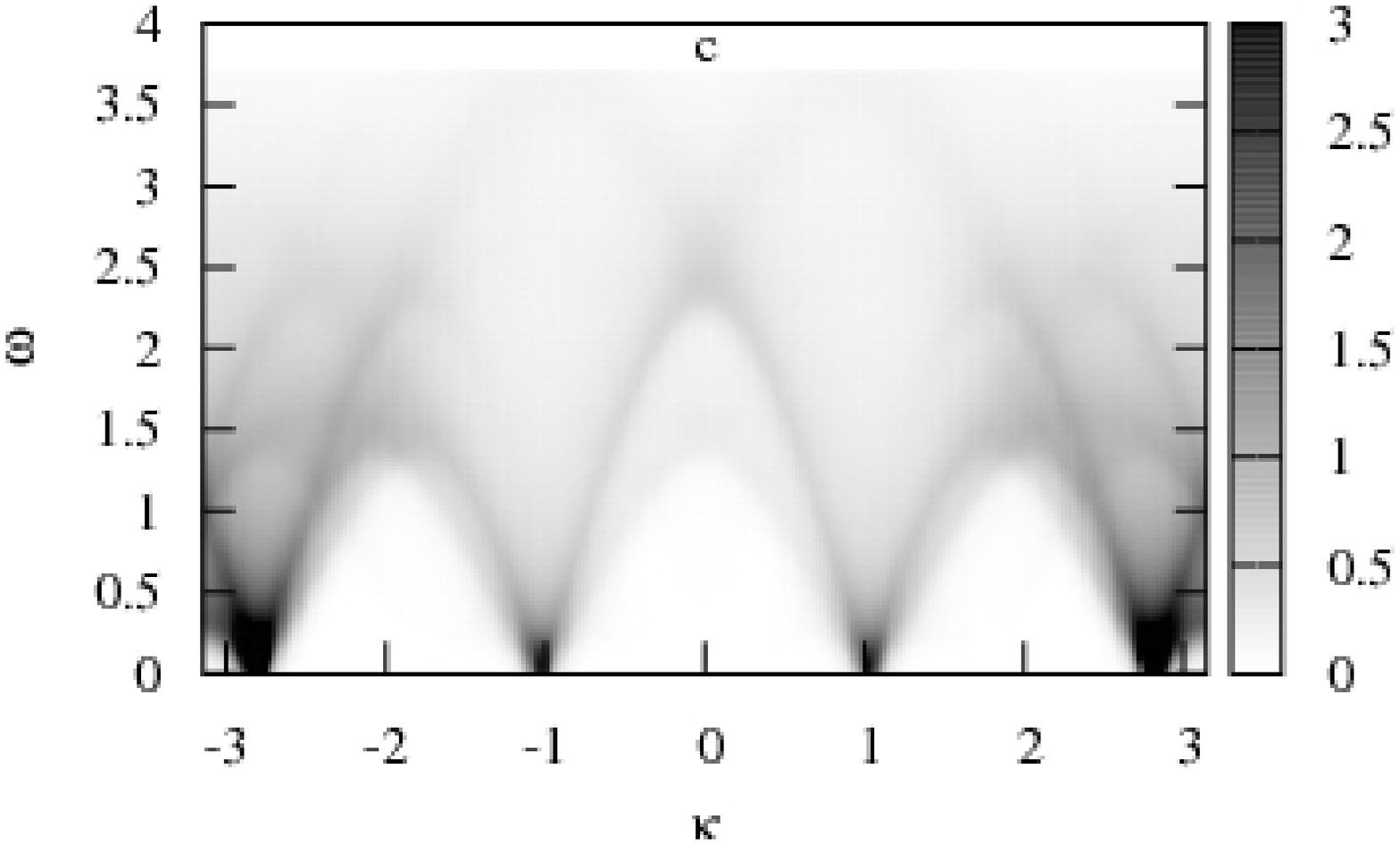, width = 0.32\linewidth}\\
\epsfig{file = 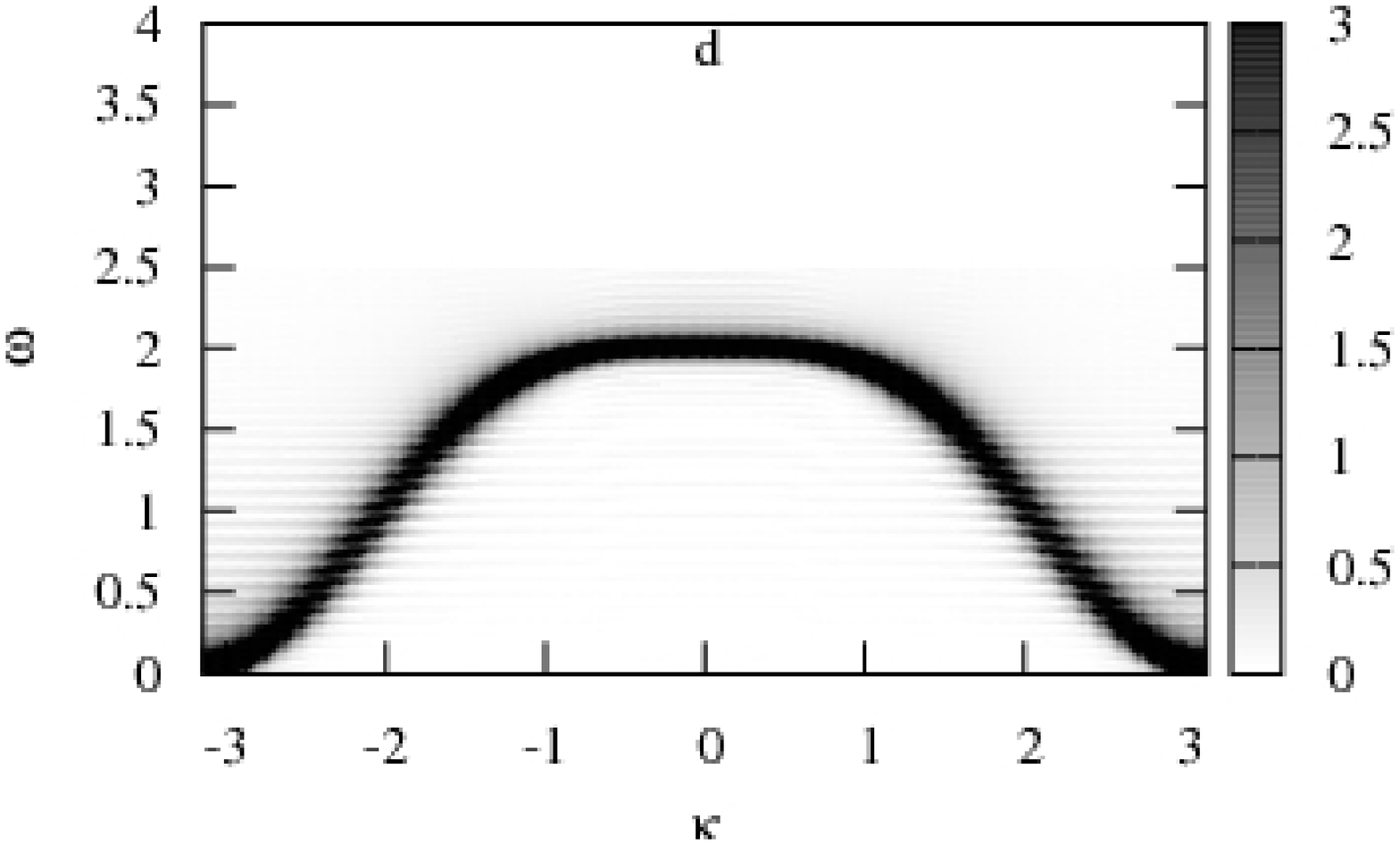, width = 0.32\linewidth}
\epsfig{file = 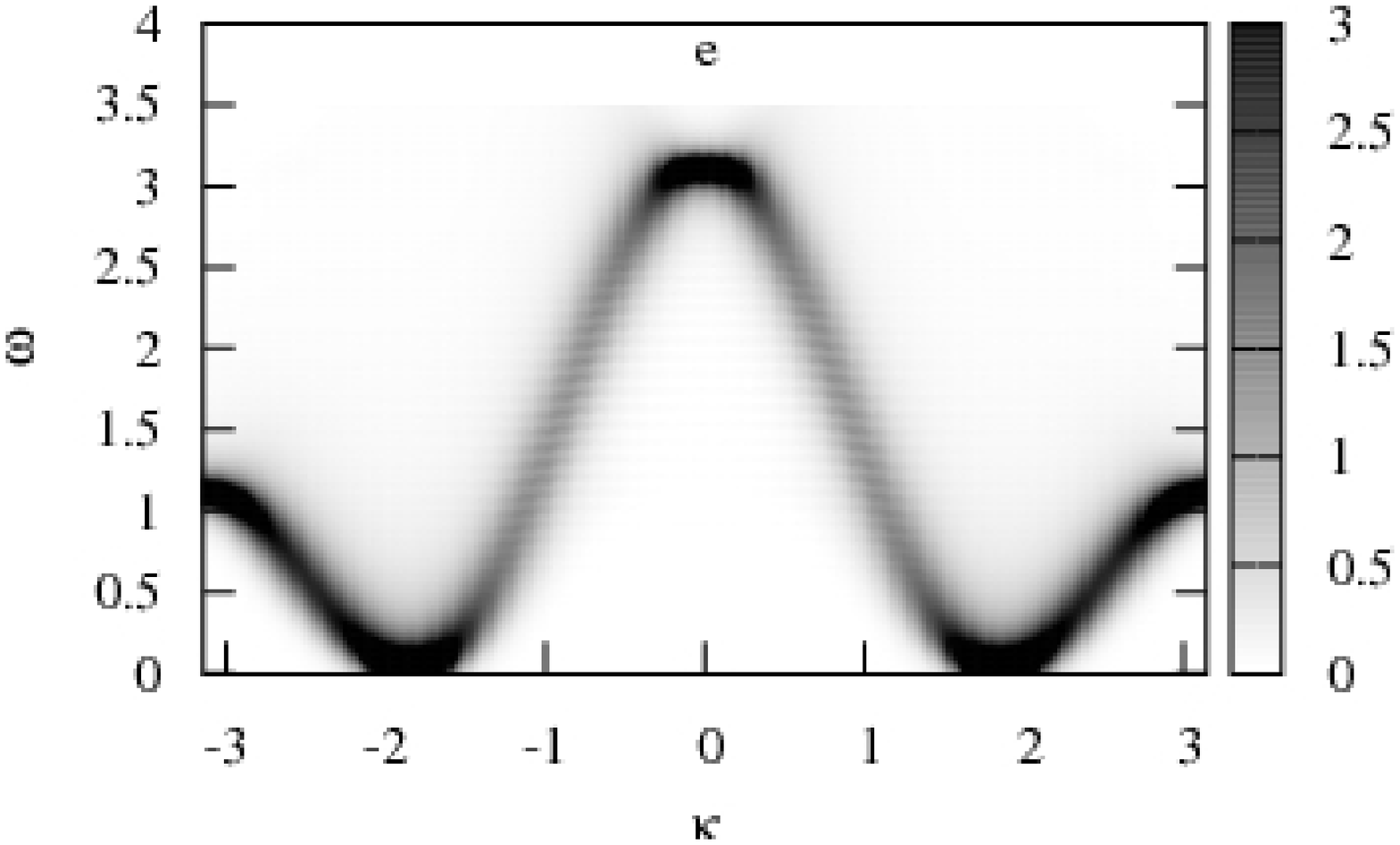, width = 0.32\linewidth}
\epsfig{file = 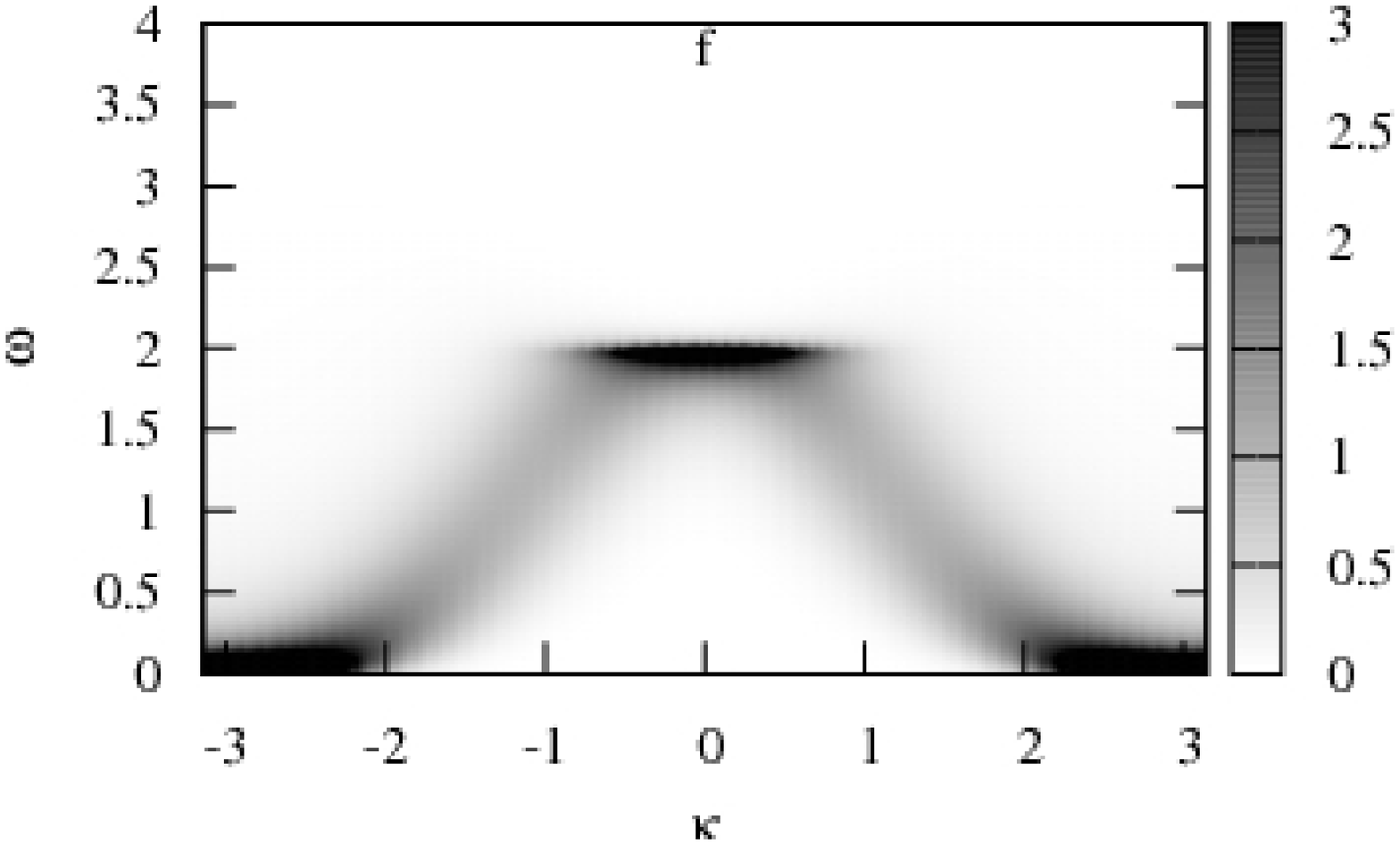, width = 0.32\linewidth}\\
\epsfig{file = 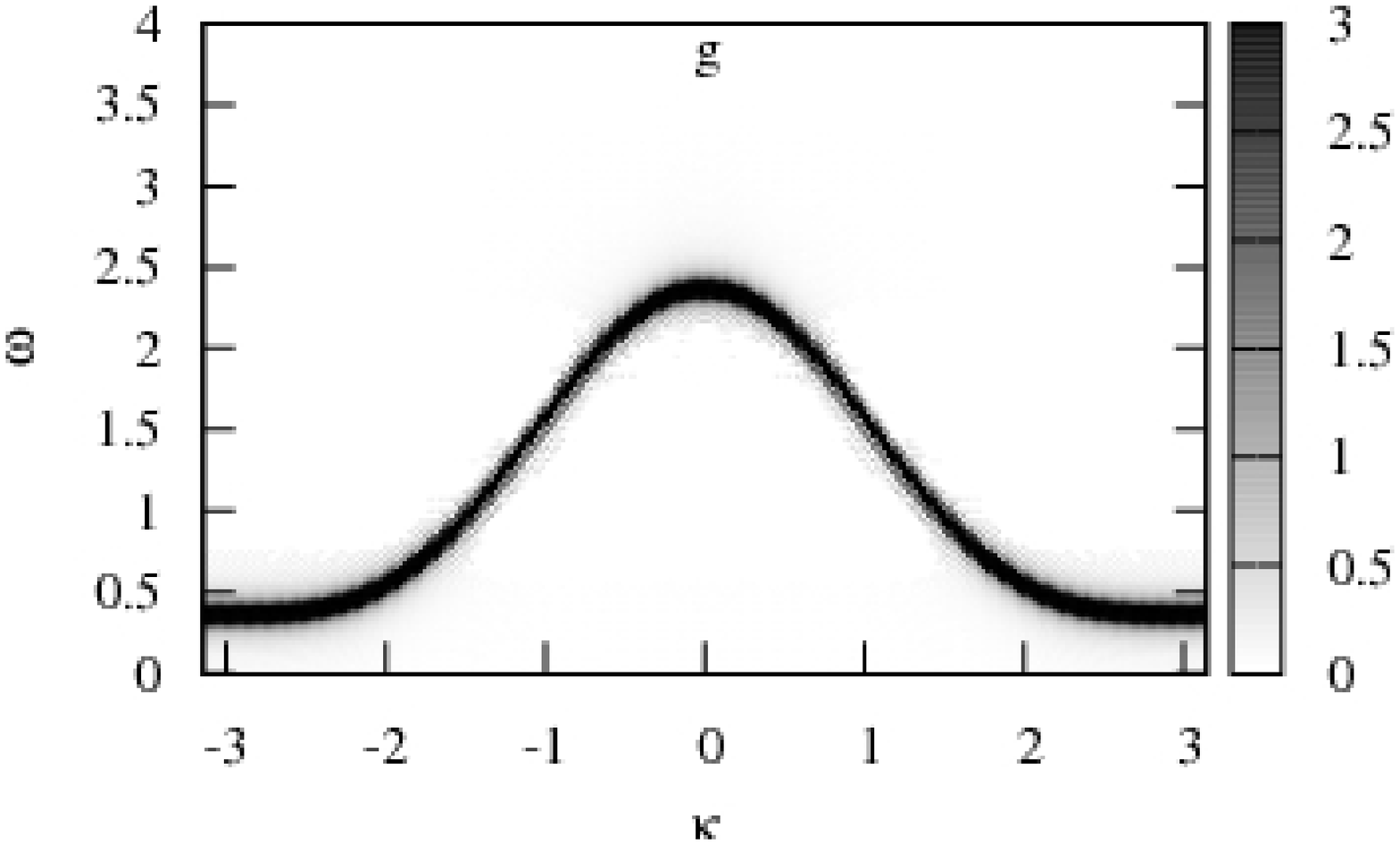, width = 0.32\linewidth}
\epsfig{file = 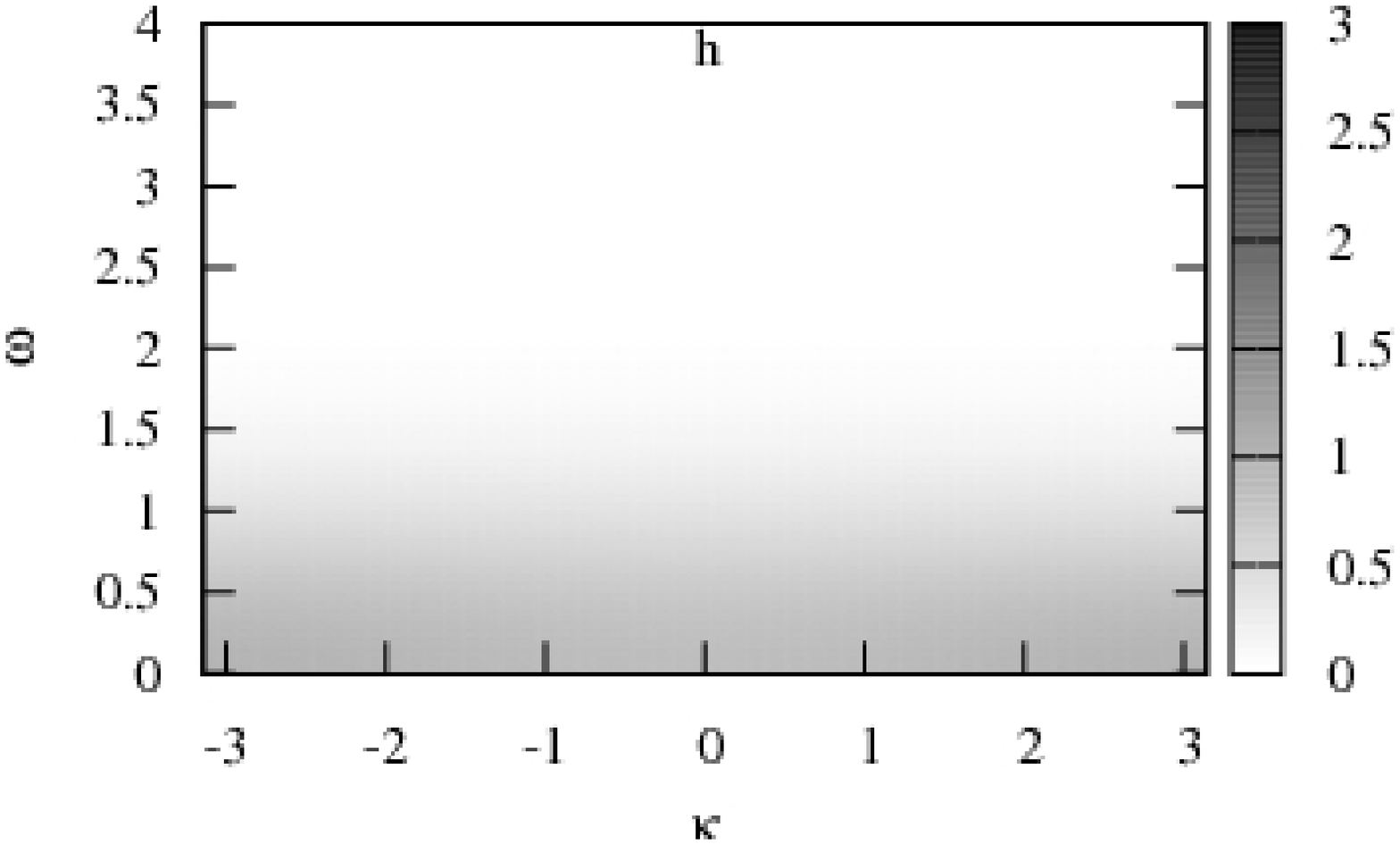, width = 0.32\linewidth}
\epsfig{file = 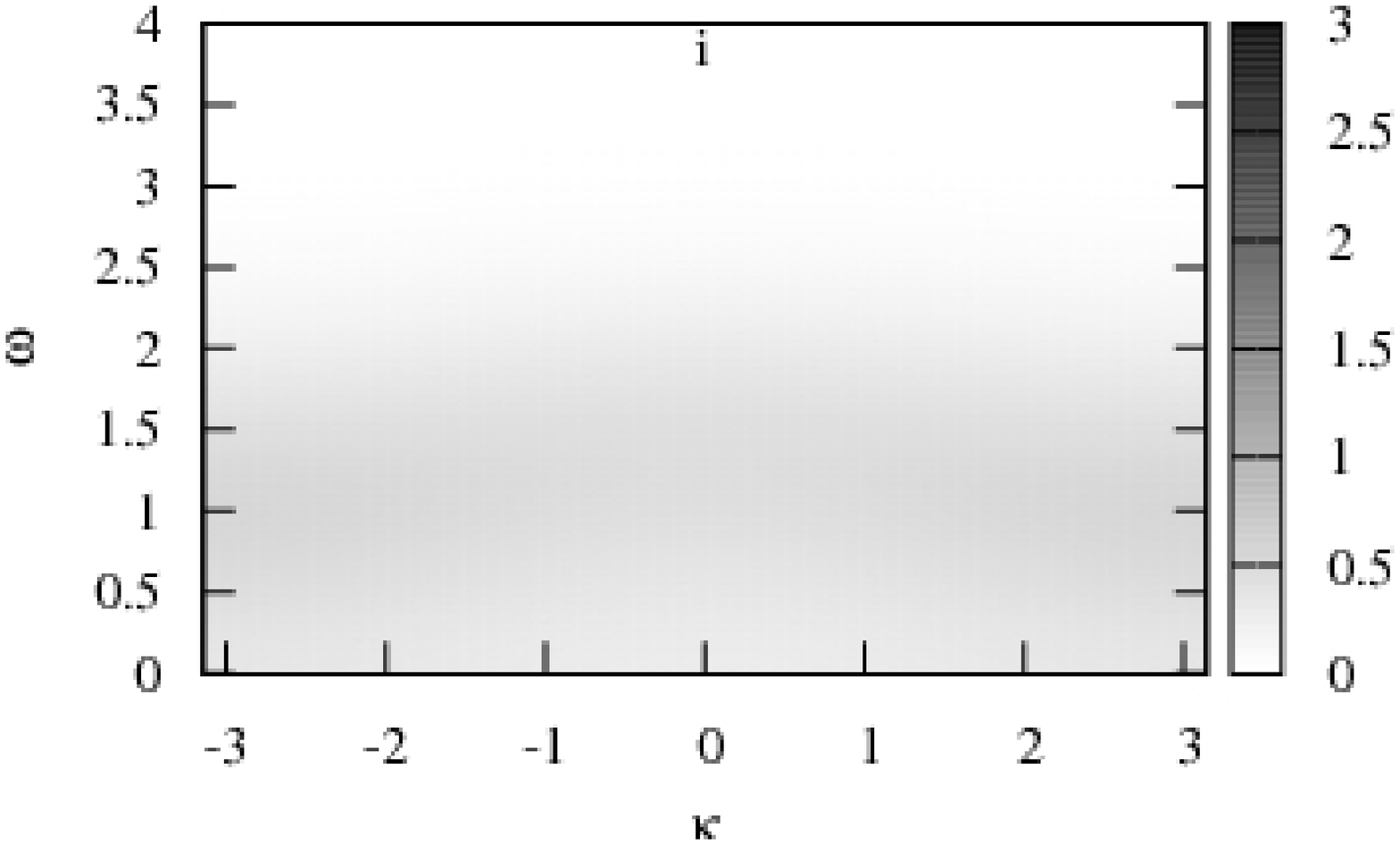, width = 0.32\linewidth}
\caption{$S_{xx}(\kappa,\omega)$
for the model (\ref{1.01}) with $J=1$, $D=E=0$,
$K=0.5$, $\Omega=0$ (panel a),
$K=2$, $\Omega=0$ (panel b),
$K=2.5$, $\Omega=0$ (panel c),
$K=0.5$, $\Omega=1.25$ (panel d),
$K=2$, $\Omega=-1.125$ (panel e),
$K=0.5$, $\Omega=-0.75$ (panel f),
$K=0.5$, $\Omega=-1.125$ 
(for this set of parameters we assume $\epsilon=0.02$, see the main text) 
(panel g)
at the low temperature $\beta=20$.
The panels h and i
correspond to high-temperature limit, $\beta=0.1$,
$K=0.5$
and $\Omega=0$ (h) or $\Omega=-1.125$ (i).}
\label{fig10}
\end{figure}

We turn next to the case of arbitrary values of temperature and the Hamiltonian parameters.
In this case we calculate the $xx$ dynamic structure factor numerically.
The numerical approach for calculating dynamic quantities was explained in detail earlier \cite{dk,dks}. 
To calculate $\langle s_j^x(t)s_{j+n}^x\rangle$ 
we express the spin operators $s^x$ entering this quantity 
in terms of the Fermi operators $c_\kappa$, $c^{\dagger}_\kappa$ according to (\ref{1.04}), (\ref{1.06}) 
obtaining as a result an average of a product of large number of Fermi operators 
attached not only to the sites $j$ and $j+n$ 
but to two strings of sites extending to the site $j=1$. 
We apply the Wick-Bloch-de Dominicis theorem 
and present the result as the Pfaffian of the $2(2j+n-1)\times 2(2j+n-1)$ antisymmetric matrix 
constructed from the known elementary contractions 
(only these quantities are influenced by the existence of three-site interactions). 
Finally, 
we evaluate numerically the Pfaffians 
obtaining as a result the desired $xx$ time-dependent spin correlation function. 
To get $S_{xx}(\kappa,\omega)$ (\ref{1.08}) 
we perform numerically the integration over time $t$ and then the summation over $n$. 
Typically we take $N=400$, 
assume $j=41$, 
calculate $\langle s_j^x(t)s_{j+n}^x\rangle$ 
for $n$ up to $n_{\max}=100$ in the time range up to $t_{\max}=100$.
(However, for large $\Omega$ we assume $j=81$, $n_{\max}=100$, $t_{\max}=200$
whereas for $\beta=0.1$ it is sufficient to take $j=41$, $n_{\max}=50$, $t_{\max}=50$.) 
In the low-temperature strong-field regime 
the $xx$ time-dependent spin correlation functions display long-time oscillations 
which lead to evident problems with integrating over time;
therefore, in this case 
(in fact, only for the set of parameters corresponding to panel g in Fig. \ref{fig10})
we introduce under the integral in Eq. (\ref{1.08}) 
an auxiliary damping factor $\exp(-\epsilon \vert t\vert)$, where $\epsilon$ is a small positive number.
We examine in detail different types of finite size effects \cite{dk,dks} 
to be sure that our results for $S_{xx}(\kappa,\omega)$ refer to the thermodynamic limit. 
In Fig. \ref{fig10} 
we demonstrate the results for the $xx$ dynamic structure factors 
for a set of parameters which is in correspondence with the one used in Fig. \ref{fig02} 
(panels a -- f) 
and the analytical predictions (\ref{4.06}), (\ref{4.07})
(panels g, h, i).

We start to discuss the results obtained for $S_{xx}(\kappa,\omega)$ from the case of low temperatures.
Although $S_{xx}(\kappa,\omega)$ is a many-fermion quantity 
and therefore is not restricted to a certain region in the $\kappa$ -- $\omega$ plane 
it is mostly concentrated along certain lines in the $\kappa$ -- $\omega$ plane 
which roughly correspond to the characteristic lines of the two-fermion excitation continuum 
discussed in Sec.~\ref{sec3}
(compare Figs. \ref{fig10}a, \ref{fig10}b, \ref{fig10}c
with Figs. \ref{fig02}a, \ref{fig02}b, \ref{fig02}c
(and Figs. \ref{fig06}a, \ref{fig06}d, \ref{fig06}g)
as well as Figs. \ref{fig10}d, \ref{fig10}e, \ref{fig10}f
with Figs. \ref{fig02}d, \ref{fig02}e, \ref{fig02}f).
For example, 
for $J=1$, $K=0.5$, $\Omega=0$ (panel a in Fig. \ref{fig10})
$S_{xx}(\kappa,\omega)$ is accumulated along the three lines in the $\kappa$ -- $\omega$ plane,
$\Omega_{-}(\kappa)$ (\ref{3.04}),
$\omega_{-}^{\pm}(\kappa)$ (\ref{3.08}),
shifted along $\kappa$-axis by $\pi$.
Although the important role of the two-fermion excitations 
for the low-temperature many-fermion dynamic quantities 
like $S_{xx}(\kappa,\omega)$ 
was noted several times earlier,
we still do not have a simple explanation for that fact.
On the other hand,
the panels d, e, f and g in Fig. \ref{fig10} demonstrate the development towards 
the zero-temperature strong-field result (\ref{4.07}).

As temperature increases the role of the two-fermion excitations diminishes
and in the high-temperature limit 
many-fermion excitations produce the $\kappa$-independent Gaussian decay (\ref{4.06})
(compare Figs. \ref{fig10}h, \ref{fig10}i to Figs. \ref{fig10}a, \ref{fig10}g).

\section{Conclusions}
\label{sec5}
\setcounter{equation}{0}

In conclusion, 
we have examined several dynamic structure factors of the spin-1/2 transverse $XX$ chain 
with $(XZX+YZY)$-type three-site interactions. 
These three-site interactions essentially enrich the ground-state phase diagram of the spin model
which may show two different spin liquid phases 
(spin liquid I and spin liquid II)
in addition to the ferromagnetic phase. 
We have calculated explicitly several dynamic structure factors 
(with the transverse dynamic structure factor $S_{zz}(\kappa,\omega)$ 
and the dimer dynamic structure factor $S_{J}(\kappa,\omega)$ among them)
which are governed exclusively by two-fermion (particle-hole) excitations.
We have discussed in some detail 
the properties of the two-fermion excitation continuum 
determining its boundaries, soft modes and exponents of van Hove singularities \cite{foot}.
We have also discussed some specific features 
of different two-fermion dynamic structure factors.
Our analysis of many-fermion dynamic structure factors 
is restricted to the $xx$ dynamic structure factor $S_{xx}(\kappa,\omega)$.
For this quantity 
we have reported exact analytical results in the high-temperature and zero-temperature strong-field limits 
and precise numerical results for other sets of parameters.
The three-site interactions introduced leave a number of signatures in the dynamic quantities 
producing an extra two-fermion excitation continuum,
van Hove singularity with exponent 2/3,
or singularity of the dimer structure factor along the upper boundary of the two-fermion excitation continuum.
In the presence of three-site interactions the symmetry of the $xx$ dynamic structure factor 
in the zero-temperature strong-field regime with respect to the change $\Omega\to -\Omega$ is broken.

We emphasize that the advantage of the model considered is its exact solvability,
that means, in particular, the possibility to calculate various dynamic quantities accurately.
On the other hand,
although there are some examples of real-life systems 
which can be modeled as spin-1/2 $XX$ chains
(see, e.g., Ref. \onlinecite{exp}),
the three-site interactions introduced are of a rather special kind,
however,
the reported results may serve to test other (approximate) techniques 
used to study more realistic models,
e.g., with next-nearest neighbor interactions or with four-site interactions.
Moreover, 
our results on dynamics may be used for discussing the effects 
of stationary energy fluxes in quantum spin chains.
Thus in Ref. \onlinecite{antal}c
the transverse ($zz$) dynamic structure factor for a model with $D=K=0$ was discussed  
in relation to possible experimental observation of energy-current carrying states 
in quantum spin chain compounds.

\section*{Acknowledgments}

This research was supported by a NATO collaborative linkage grant,
reference number CBP.NUKR.CLG 982540,
project ``Dynamic Probes of Low-Dimensional Quantum Magnets''.
J.~S. acknowledges the kind hospitality of ICMP, L'viv, in 2007
where part of this work was done.
The authors thank U.~L\"{o}w for critical reading of the manuscript.

\appendix
\section*{Appendix: Time-dependent $xx$ spin correlations in the zero-temperature strong-field regime}

\label{seca}
\renewcommand{\theequation}{A.\arabic{equation}}
\setcounter{equation}{0}

We can rewrite Eq. (\ref{4.04}) ($\Omega>0$ \cite{negative}) as follows
\begin{equation}
\label{a.01}
4 \langle s^x_j(t)s^x_{j+n}\rangle 
=\exp\left(-{\rm{i}}\left(\Omega+\frac{K}{2}\right)t\right) {\cal{L}}_n
\end{equation}
with
\begin{eqnarray}
\label{a.02}
{\cal{L}}_n
\equiv\frac{1}{\pi}\int_0^{\pi}{\rm{d}}\kappa 
\cos(\kappa n)
\exp\left(-{\rm{i}}t\left(J\cos\kappa - K\cos^2\kappa\right)\right). 
\end{eqnarray}
The function ${\cal{L}}_n$ introduced in (\ref{a.02}) for even $n$ 
can be expressed in terms of the function $\Phi_3(\beta,\gamma,x,y)$ 
(see section 5.7.1 in Ref. \onlinecite{be1})
\begin{eqnarray}
\label{a.03}
{\cal{L}}_{0}=\exp({\rm{i}}Kt) 
\Phi_3\left(\frac{1}{2},1;-{\rm{i}}Kt,\frac{J^2t^2}{4}\right),
\nonumber\\
{\cal{L}}_{2m}=\exp({\rm{i}}Kt) 
\sum_{l=0}^m(-1)^l\frac{m\Gamma(m+l)}{\Gamma(m-l+1)\Gamma^2(l+1)}
\Phi_3\left(l+\frac{1}{2},l+1;-{\rm{i}}Kt,\frac{J^2t^2}{4}\right);
\end{eqnarray}
here $\Gamma(n)$ is the gamma function.

We notice that in the case $K=0$
\begin{eqnarray}
\label{a.04}
{\cal{L}}_n=(-{\rm{i}})^{n}J_n(Jt),
\end{eqnarray}
whereas in the case $J=0$
\begin{eqnarray}
\label{a.05}
{\cal{L}}_n
=
\left\{
\begin{array}{ll}
\exp\left({\rm{i}}\frac{Kt}{2}\right) {\rm{i}}^m J_n\left(\frac{Kt}{2}\right), & n=2m, \\
0,                                                                             & n=2m+1;
\end{array}
\right.
\end{eqnarray}
here $J_n(z)$ is the Bessel function of the first kind.

Finally,
we notice that the long-time asymptotics for $\langle s^x_j(t)s^x_{j+n}\rangle$
in the zero-temperature strong-field regime
can be calculated accurately
using the stationary phase method \cite{fedoryuk}.
For
$\vert K/J\vert<1/2$
we have
\begin{eqnarray}
\label{a.06}
{\cal{L}}_n
\stackrel{t\to\infty}{\approx}
\frac{1}{\sqrt{2\pi\vert J\vert t}}
\exp\left({\rm{i}} K t\right)
\nonumber\\ 
\times
\left( 
\frac{\exp\left(-{\rm{i}}Jt + {\rm{i}}\frac{\pi}{4}{\rm{sgn}}(J)\right)}
{\sqrt{\left\vert 1 - \frac{2K}{J}\right\vert}}
+
\frac{\exp\left({\rm{i}}Jt - {\rm{i}}\frac{\pi}{4}{\rm{sgn}}(J) + {\rm{i}}\pi n\right)}
{\sqrt{\left\vert 1\!+\!\frac{2K}{J}\right\vert}}
\right).
\end{eqnarray}
For
$\vert K/J\vert>1/2$
we have
\begin{eqnarray}
\label{a.07}
{\cal{L}}_n
\stackrel{t\to\infty}{\approx}
\frac{1}{\sqrt{2\pi\vert J\vert t}}
\exp\left({\rm{i}} K t\right) 
\nonumber\\ 
\times
\left( 
\frac{\exp\left(-{\rm{i}}Jt + {\rm{i}}\frac{\pi}{4}{\rm{sgn}}(J-2K)\right)}
{\sqrt{\left\vert 1 - \frac{2K}{J}\right\vert}} 
+\frac{\exp\left({\rm{i}}Jt - {\rm{i}}\frac{\pi}{4}{\rm{sgn}}(J+2K) + {\rm{i}}\pi n\right)}
{\sqrt{\left\vert 1\!+\!\frac{2K}{J}\right\vert}} 
\right. 
\nonumber\\
\left. 
+2\frac{\exp\left(-{\rm{i}}\left(K+\frac{J^2}{4K}\right)t
+
{\rm{i}}\frac{\pi}{4}{\rm{sgn}}\left(2K - \frac{J^2}{2K}\right)\right)}
{\sqrt{\left\vert\frac{2K}{J}-\frac{J}{2K}\right\vert}}
\cos\left(n\arccos\frac{J}{2K}\right)
\right).
\end{eqnarray}
For
$\vert K/J\vert=1/2$
we have
\begin{eqnarray}
\label{a.08}
{\cal{L}}_n
\stackrel{t\to\infty}{\approx}
\left(-1\right)^{\frac{n}{2}\left(1-{\rm{sgn}}(JK)\right)}
\left(
\frac{\Gamma\left(\frac{1}{4}\right)}{2\pi\left(2\vert J\vert t\right)^{\frac{1}{4}}} 
\exp\left(
{\rm{i}}\,{\rm{sgn}}(JK)\left(-\frac{1}{2}Jt + \frac{\pi}{8}{\rm{sgn}}(J)\right)
\right)
\right.
\nonumber\\
\left.
+\frac{1}{2\sqrt{\pi\vert J\vert t}}
\exp\left(
{\rm{i}}\,{\rm{sgn}}(JK)\left(\frac{3}{2}Jt - \frac{\pi}{4}{\rm{sgn}}(J)\right) + {\rm{i}}\pi n
\right)
\right).
\end{eqnarray}
The long-time asymptotic behavior (\ref{a.06}), (\ref{a.07}), (\ref{a.08}) 
may emerge already at relatively short times 
as can be seen in Fig. \ref{fig09}.


\begin{thebibliography}{99}
\bibitem{lsm_k}
E.~Lieb, T.~Schultz, and D.~Mattis,
Ann. Phys. (N.Y.) {\bf 16}, 407 (1961);\\
S.~Katsura,
Phys. Rev. {\bf 127}, 1508 (1962);
{\bf 129}, 2835 (1963).

\bibitem{spin_from_hubb}
K.~Yosida,
Theory of Magnetism 
(Springer-Verlag, Berlin, 1996);\\
A.~H.~MacDonald, S.~M.~Girvin, and D.~Yoshioka,
Phys. Rev. B {\bf 37}, 9753 (1988);\\
A.~H.~MacDonald, S.~M.~Girvin, and D.~Yoshioka,
Phys. Rev. B {\bf 41}, 2565 (1990);\\
A.~L.~Chernyshev, D.~Galanakis, P.~Phillips, A.~V.~Rozhkov, and A.-M.~S.~Tremblay,
Phys. Rev. B {\bf 70}, 235111 (2004);\\
A.~Reischl, E.~M\"{u}ller-Hartmann, and G.~S.~Uhrig,
Phys. Rev. B {\bf 70}, 245124 (2004).

\bibitem{antal}
T.~Antal, Z.~R\'{a}cz, A.~R\'{a}kos, and G.~M.~Sch\"utz,
Phys. Rev. E {\bf 57}, 5184 (1998);\\
T.~Antal, Z.~R\'{a}cz, A.~R\'{a}kos, and G.~M.~Sch\"utz,
Phys. Rev. E {\bf 59}, 4912 (1999);\\
Z.~R\'{a}cz, J. Stat. Phys. {\bf 101}, 273 (2000).

\bibitem{ogata}
Y.~Ogata,
Phys. Rev. E {\bf 66}, 016135 (2002).

\bibitem{eisler}
V.~Eisler and Z.~Zimbor\'{a}s,
Phys. Rev. A {\bf 71}, 042318 (2005).

\bibitem{suzuki}
M.~Suzuki,
Phys. Lett. A {\bf 34}, 338 (1971);\\
M.~Suzuki,
Prog. Theor. Phys. {\bf 46}, 1337 (1971).

\bibitem{tsvelik}
A.~M.~Tsvelik,
Phys. Rev. B {\bf 42}, 779 (1990).

\bibitem{frahm}
H.~Frahm,
J. Phys. A {\bf 25}, 1417 (1992).

\bibitem{gr}
D.~Gottlieb and J.~R\"{o}ssler,
Phys. Rev. B {\bf 60}, 9232 (1999).

\bibitem{drd}
O.~Derzhko, J.~Richter, and V.~Derzhko,
Ann. Phys. (Leipzig) {\bf 8}, SI-49 (1999);
arXiv:cond-mat/9908425.

\bibitem{tj}
I.~Titvinidze and G.~I.~Japaridze,
Eur. Phys. J. B {\bf 32}, 383 (2003).

\bibitem{lou1}
P.~Lou, W.-C.~Wu, and M.-C.~Chang,
Phys. Rev. B {\bf 70}, 064405 (2004).

\bibitem{lou2}
P.~Lou,
phys. stat. sol. (b) {\bf 241}, 1343 (2004).

\bibitem{lou3}
P.~Lou,
Phys. Rev. B {\bf 72}, 064435 (2005).

\bibitem{lou4}
P.~Lou,
phys. stat. sol. (b) {\bf 242}, 3209 (2005).

\bibitem{pachos}
J.~K.~Pachos,
International Journal of Quantum Information {\bf 4}, 541 (2006);
arXiv:quant-ph/0505225.

\bibitem{yang}
M.-F.~Yang,
Phys. Rev. A {\bf 71}, 030302(R) (2005).

\bibitem{zhang}
J.~Zhang, X.~Peng, and D.~Suter,
Phys. Rev. A {\bf 73}, 062325 (2006).

\bibitem{lou5}
P.~Lou and J.~Y.~Lee,
Phys. Rev. B {\bf 74}, 134402 (2006).

\bibitem{yu}
C.-s.~Yu, H.-s.~Song, and H.-t.~Cui,
arXiv:quant-ph/0611011.

\bibitem{z1}
A.~A.~Zvyagin and G.~A.~Skorobagat'ko,
Phys. Rev. B {\bf 73}, 024427 (2006).

\bibitem{z2}
A.~A.~Zvyagin,
Phys. Rev. B {\bf 73}, 104414 (2006).

\bibitem{taylor}
J.~H.~Taylor and G.~M\"{u}ller,
Physica A {\bf 130}, 1 (1985)
and references therein.

\bibitem{dks}
O.~Derzhko, T.~Krokhmalskii, and J.~Stolze,
J. Phys. A {\bf 33}, 3063 (2000).

\bibitem{dksm}
O.~Derzhko, T.~Krokhmalskii, J.~Stolze, and G.~M\"{u}ller,
Phys. Rev. B {\bf 71}, 104432 (2005).

\bibitem{z3}
A.~A.~Zvyagin,
Phys. Rev. B {\bf 72}, 064419 (2005).

\bibitem{perk}
J.~H.~H.~Perk and H.~W.~Capel,
Phys. Lett. A {\bf 58}, 115 (1976);\\
M.~Oshikawa and I.~Affleck,
Phys. Rev. Lett. {\bf 79}, 2883 (1997);\\
O.~Derzhko, J.~Richter, and O.~Zaburannyi,
J. Phys.: Condens. Matter {\bf 12}, 8661 (2000);\\
D.~N.~Aristov and S.~V.~Maleyev,
Phys. Rev. B {\bf 62}, R751 (2000);\\
M.~Bocquet, F.~H.~L.~Essler, A.~M.~Tsvelik, and A.~O.~Gogolin,
Phys. Rev. B {\bf 64}, 094425 (2001).

\bibitem{sachdev}
S.~Sachdev,
Quantum Phase Transitions
(Cambridge University Press, Cambridge, UK, 1999).

\bibitem{hubbard}
S.~Daul and R.~M.~Noack,
Phys. Rev. B {\bf 58}, 2635 (1998);\\
S.~Daul and R.~M.~Noack,
Phys. Rev. B {\bf 61}, 1646 (2000);\\
G.~I.~Japaridze, R.~M.~Noack, D.~Baeriswyl, and L.~Tincani,
Phys. Rev. B {\bf 76}, 115118 (2007);\\
S.~Nishimoto, K.~Sano, and Y.~Ohta,
arXiv:0710.2274v1 [cond-mat.str-el].

\bibitem{zubarev}
D.~N.~Zubarev,
Njeravnovjesnaja Statistitchjeskaja Tjermodinamika
(Nauka, Moskva, 1971)
(in Russian);
D.~N.~Zubarev,
Nonequilibrium Statistical Thermodynamics
(Consultants Bureau, New York, 1974).

\bibitem{kdsv}
T.~Krokhmalskii, O.~Derzhko, J.~Stolze, and T.~Verkholyak,
Acta Physica Polonica A {\bf 113}, 437 (2008).

\bibitem{niemejer}
Th.~Niemejer,
Physica {\bf 36}, 377 (1967).

\bibitem{dls}
O.~V.~Derzhko, R.~R.~Levitskii, and S.~I.~Sorokov,
Ukrainian Journal of Physics {\bf 35}, 1421 (1990) (in Ukrainian).

\bibitem{florencio}
J.~Florencio,~Jr. and M.~H.~Lee,
Phys. Rev. {\bf 35}, 1835 (1987).

\bibitem{stolze}
U.~Brandt and J.~Stolze,
Z. Phys. B {\bf 64}, 327 (1986);\\
J.~M.~R.~Roldan, B.~M.~McCoy, and J.~H.~H.~Perk,
Physica A {\bf 136}, 255 (1986);\\
J.~Stolze, V.~S.~Viswanath, and G.~M\"{u}ller,
Z. Phys. B {\bf 89}, 45 (1992);\\
M.~B\"{o}hm, H.~Leschke, M.~Henneke, V.~S.~Viswanath, J.~Stolze, and G.~M\"{u}ller,
Phys. Rev. B {\bf 49}, 417 (1994);\\
M.~B\"{o}hm, V.~S.~Viswanath, J.~Stolze, and G.~M\"{u}ller,
Phys. Rev. B {\bf 49}, 15669 (1994).

\bibitem{inf_temp}
U.~Brandt and K.~Jacoby,
Z. Phys. B {\bf 25}, 181 (1976);\\
H.~W.~Capel and J.~H.~H.~Perk,
Physica A {\bf 87}, 211 (1977).

\bibitem{perk_capel_1978}
J.~H.~H.~Perk and H.~W.~Capel, 
Physica A {\bf 92}, 163 (1978).

\bibitem{zero_temp}
H.~B.~Cruz and L.~L.~Gon\c{c}alves,
J. Phys. C {\bf 14}, 2785 (1981). 

\bibitem{be2}
H.~Bateman, A.~Erd\'{e}lyi, 
Higher Transcendental Functions, vol.~2 
(McGraw-Hill, New York, 1953).

\bibitem{dk}
O.~Derzhko and T.~Krokhmalskii,
Phys. Rev. B {\bf 56}, 11659 (1997);\\
O.~Derzhko and T.~Krokhmalskii,
phys. stat. sol. (b) {\bf 208}, 221 (1998).

\bibitem{foot}
Modification of critical exponents 
due to competition of nearest-neighbor and next-nearest-neighbor interactions 
has also been observed recently in:
C.~Trippe and A.~Kl\"{u}mper,
Fizika Nizkikh Temperatur (Kharkiv) {\bf 33}, 1213 (2007);
Low Temperature Physics {\bf 33}, 920 (2007).

\bibitem{exp}
M.~Kenzelmann, R.~Coldea, A.~A.~Tennant, D.~Visser, M.~Hofmann, P.~Smeibidl, and Z.~Tylczynski,
Phys. Rev. B {\bf 65}, 144432 (2002).

\bibitem{negative}
The results for $\Omega<0$ can be obtained by symmetry 
after replacing $\Omega$, $K$ to $-\Omega$, $-K$.

\bibitem{be1}
H.~Bateman, A.~Erd\'{e}lyi, 
Higher Transcendental Functions, vol.~1 
(McGraw-Hill, New York, 1953).

\bibitem{fedoryuk}
M.~V.~Fedoryuk,
Mjetod Pjerjevala
(Nauka, Moskva, 1977)
(in Russian).

\end{thebibliography}
\end{document}